\documentclass[times,twocolumn,final]{elsarticle} 
\usepackage{medima}
\usepackage{framed,multirow}

\usepackage{latexsym}

\usepackage{url}
\usepackage{xcolor}
\usepackage{cite}

\usepackage{amsmath,amssymb,amsfonts}
\usepackage{algorithmic}
\usepackage{graphicx}
\usepackage{textcomp}

\usepackage{url}
\usepackage{cite}

\usepackage{booktabs}
\usepackage{diagbox}
\usepackage{tabularx}
\usepackage{setspace}
\usepackage{color}
\usepackage[colorlinks,linkcolor=blue]{hyperref}
\usepackage{cleveref}
\usepackage{multirow}
\usepackage{tablefootnote}
\usepackage{threeparttable}
\usepackage{url}
\usepackage{caption}
\usepackage{bm}

\usepackage{longtable}

\definecolor{newcolor}{rgb}{.8,.349,.1}

\journal{Medical Image Analysis}

\newcommand{\ourdata}{COSMOS 1050K}

\begin{document}

\verso{Yuhao Huang \textit{et~al.}}

\begin{frontmatter}

\title{Segment Anything Model for Medical Images?}%


\author[1,2,3]{Yuhao Huang\fnref{fn1}}
\author[1,2,3]{Xin Yang\fnref{fn1}}
\author[1,2,3]{Lian Liu}
\author[1,2,3]{Han Zhou}
\author[1,2,3]{Ao Chang}
\author[1,2,3]{Xinrui Zhou}
\author[1,2,3]{Rusi Chen}
\author[1,2,3]{Junxuan Yu}
\author[1,2,3]{Jiongquan Chen}
\author[1,2,3]{Chaoyu Chen}
\author[1,2,3]{Sijing Liu}
\author[4]{Haozhe Chi}
\author[5]{Xindi Hu}
\author[6]{Kejuan Yue}
\author[7]{Lei Li}
\author[7]{Vicente Grau}
\author[8]{Deng-Ping Fan}
\author[9,10]{Fajin Dong\corref{cor1}}
\ead{dongfajin@szhospital.com}
\author[1,2,3]{Dong Ni\corref{cor1}}
\ead{nidong@szu.edu.cn}

\cortext[cor1]{Corresponding author. {$^{1}$}Authors contributed equally.}

\address[1]{National-Regional Key Technology Engineering Laboratory for Medical Ultrasound, School of Biomedical Engineering, Shenzhen University Medical School, Shenzhen University, Shenzhen, China}
\address[2]{Medical UltraSound Image Computing (MUSIC) Lab, Shenzhen University, Shenzhen, China}
\address[3]{Marshall Laboratory of Biomedical Engineering, Shenzhen University, Shenzhen, China}
\address[4]{Zhejiang University, Zhejiang, China}
\address[5]{Shenzhen RayShape Medical Technology Co., Ltd, Shenzhen, China}
\address[6]{Hunan First Normal University, Changsha, China}
\address[7]{Department of Engineering Science, University of Oxford, Oxford, UK}
\address[8]{Computer Vision Lab (CVL), ETH Zurich, Zurich, Switzerland}
\address[9]{Ultrasound Department, the Second Clinical Medical College, Jinan University, China}
\address[10]{First Affiliated Hospital, Southern University of Science and Technology, Shenzhen People’s Hospital, Shenzhen, China}

\received{*****}
\finalform{*****}
\accepted{*****}
\availableonline{*****}
\communicated{S. Sarkar}

\begin{abstract}
The Segment Anything Model (SAM) is the first foundation model for general image segmentation.
It has achieved impressive results on various natural image segmentation tasks.
However, medical image segmentation (MIS) is more challenging because of the complex modalities, fine anatomical structures, uncertain and complex object boundaries, and wide-range object scales.
To fully validate SAM's performance on medical data, we collected and sorted 53 open-source datasets and built a large medical segmentation dataset with 18 modalities, 84 objects, 125 object-modality paired targets, 1050K 2D images, and 6033K masks.
We comprehensively analyzed different models and strategies on the so-called \textit{\ourdata}~dataset. 
Our findings mainly include the following:
1) SAM showed remarkable performance in some specific objects but was unstable, imperfect, or even totally failed in other situations.
2) SAM with the large ViT-H showed better overall performance than that with the small ViT-B. 
3) SAM performed better with manual hints, especially box, than the \textit{Everything} mode.
4) SAM could help human annotation with high labeling quality and less time.
5) SAM was sensitive to the randomness in the center point and tight box prompts, and may suffer from a serious performance drop.
6) SAM performed better than interactive methods with one or a few points, but will be outpaced as the number of points increases. 
7) SAM's performance correlated to different factors, including boundary complexity, intensity differences, etc.
8) Finetuning the SAM on specific medical tasks could improve its average DICE performance by 4.39\% and 6.68\% for \textit{ViT-B} and \textit{ViT-H}, respectively.
Codes and models are available at: \url{https://github.com/yuhoo0302/Segment-Anything-Model-for-Medical-Images}. 
We hope that this comprehensive report can help researchers explore the potential of SAM applications in MIS, and guide how to appropriately use and develop SAM.
\end{abstract}


\end{frontmatter}

\section{Introduction}
The emergence of large language models such as ChatGPT\footnote{\url{https://chat.openai.com}} and GPT-4\footnote{\url{https://openai.com/research/gpt-4}} has sparked a new era in natural language processing (NLP), characterized by their remarkable zero-shot and few-shot generalization abilities. 
This progress has inspired researchers to develop similarly large-scale foundational models for computer vision (CV).
The first proposed foundational CV models have been primarily based on pre-training methods such as CLIP~\citep{radford2021learning} and ALIGN~\citep{jia2021scaling}.
CLIP can recognize and understand visual concepts and details, such as object shape, texture, and color, by associating them with their corresponding textual descriptions. This allows CLIP to perform a wide range of tasks, including image classification, object detection, and even visual question answering.
ALIGN can generate natural language descriptions of image regions, providing more detailed and interpretable results than traditional image-captioning approaches.
DALL·E~\citep{ramesh2021zero} was developed to generate images from textual descriptions.
This model was trained on a large dataset of text-image pairs that could create a wide range of images, from photorealistic objects to surreal scenes that combine multiple concepts.
However, these models have not been explicitly optimized for image segmentation, particularly medical image segmentation (MIS).

Recently, the \textit{Segment Anything Model (SAM)} was proposed as an innovative foundational model for image segmentation~\citep{kirillov2023segment}.
SAM was based on the vision transformer (ViT)~\citep{dosovitskiy2020image} model and trained on a large dataset with 11 million images containing 1 billion masks.
The biggest highlight of SAM is its good zero-shot segmentation performance for unseen datasets and tasks.
This process is driven by different prompts, \textit{e.g.,} points and boxes, for indicating the pixel-level semantics and region-level positions of the target objects.
It has been proven to be highly versatile and capable of addressing a wide range of segmentation tasks~\citep{kirillov2023segment}.

\begin{figure*}[!ht]
	\centering
	\includegraphics[width=0.9 \linewidth]{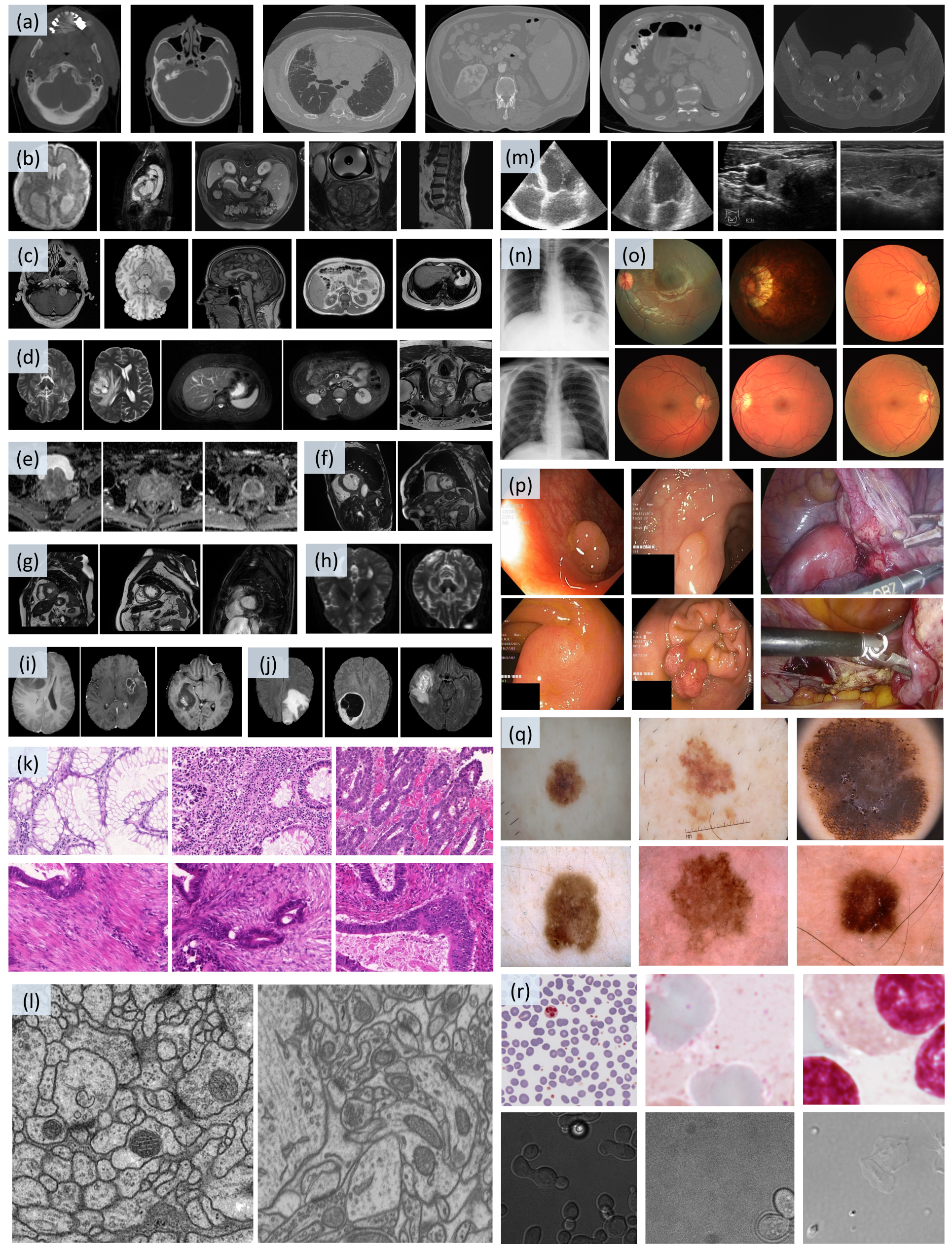}
	\caption{Our \textit{\ourdata}~dataset contains various modalities involving (a) CT, (b) MRI, (c) T1-weighted (T1W) MRI, (d) T2-weighted (T2W) MRI, (e) ADC MRI, (f) Cine-MRI, (g) CMR, (h) diffusion-weighted (DW) MRI, (i) post-contrast T1-weighted (T1-GD) MRI, (j) T2 Fluid Attenuated Inversion Recovery (T2-FLAIR) MRI, (k) Histopathology, (l) Electron Microscopy, (m) Ultrasound (US), (n) X-ray, (o) Fundus, (p) Colonoscopy, (q) Dermoscopy, and (r) Microscopy.}
	\label{fig:modal_more_data}
	\vspace{0.2mm}
\end{figure*}

\begin{figure*}[!ht]
	\centering
	\includegraphics[width=1 \linewidth]{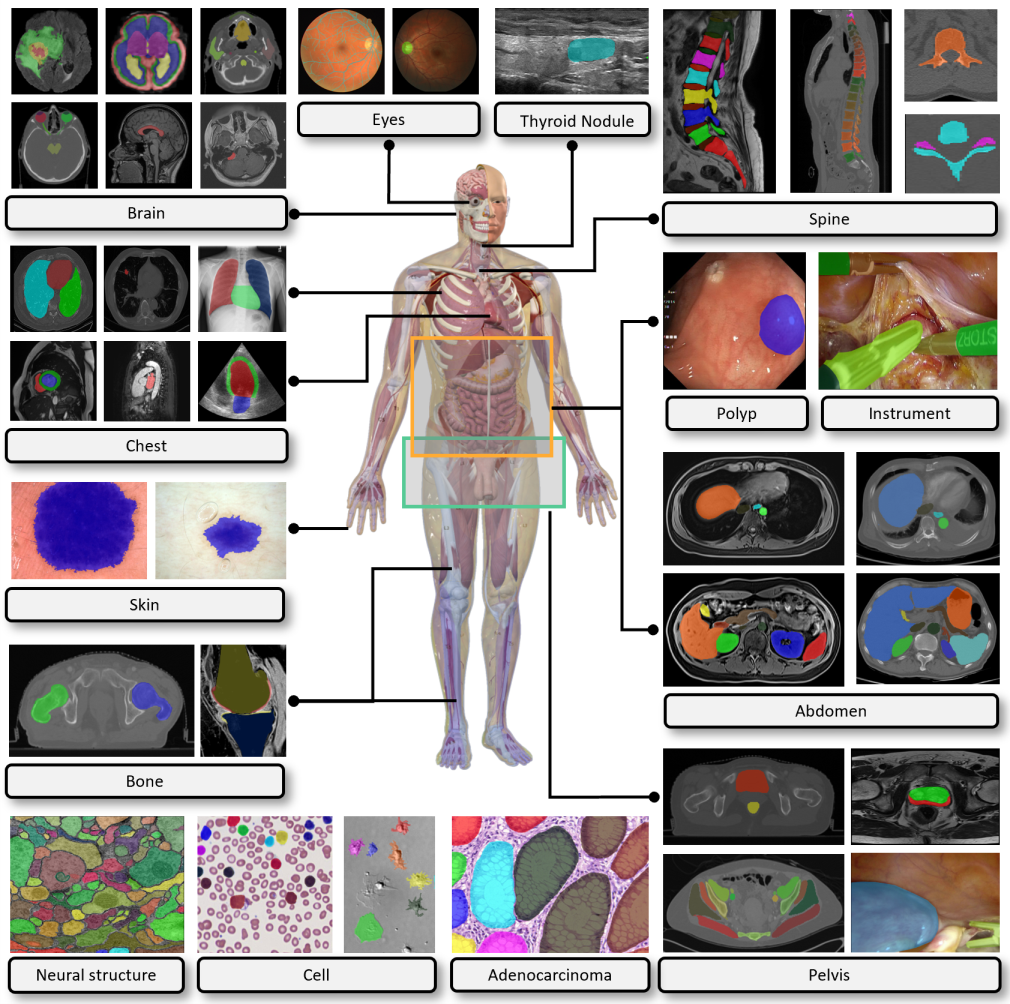}
	\caption{Our \textit{\ourdata}~dataset covers the majority of biomedical objects, for example, brain tumors, fundus vasculature, thyroid nodules, spine, lung, heart, abdominal organs and tumors, cell, polyp, and instrument.}
	\label{fig:organ_data}
\end{figure*}

Based on the pre-trained models of SAM, several papers have further studied its performance in different zero-shot segmentation scenarios.
We roughly divide them into two categories: 1) non-medical and 2) medical applications.

\subsection{SAM in Non-medical Image Applications}
Two studies focused on testing SAM's performance under the \textit{Everything} mode in segmenting the camouflaged objects~\citep{tang2023can,ji2023sam}.
Results showed that its performance was poor in these scenes, \textit{e.g.,} camouflaged animals that are visually hidden in their natural surroundings.
The authors found that SAM failed to detect concealed defects in industrial scenes~\citep{ji2023sam}.
\citet{ji2023segment} explored three testing methods of the SAM (\textit{points}, \textit{boxes}, and \textit{everything}) for various applications.
Specifically, their tasks covered natural images (salient/camouflaged/transparent object segmentation, and shadow detection), agriculture (crop segmentation and pest and leaf disease monitoring), manufacturing (anomaly and surface defect detection), and remote sensing (building and road extraction).
They concluded that, although SAM can achieve good performance in some scenarios, such as salient object segmentation and agricultural analysis, it produces poor results in other applications.
They also validated that human prompts can effectively refine the segmentation results compared to the automatic \textit{Everything} approach.

\subsection{SAM in Medical Image Analysis}
\citet{ji2023sam} assessed the SAM under the \textit{Everything} mode in segmenting the lesion regions in various anatomical structures (\textit{e.g.,} brain, lung, and liver) and modalities (computerized tomography (CT) and magnetic resonance imaging (MRI)).
The experimental results indicated that SAM was relatively proficient in segmenting organ regions with clear boundaries but may struggle to accurately identify amorphous lesion regions.
Another study then evaluated the performance of SAM in some healthcare subfields (optical disc and cup, polyp, and skin lesion segmentation) using both automatic \textit{Everything} and two manual \textit{Prompt} (points and boxes) strategies~\citep{ji2023segment}. 
The authors found that SAM required substantial human prior knowledge (\textit{i.e.,} points) to obtain relatively accurate results on these tasks. 
Otherwise, SAM resulted in wrong segmentation, especially when no prompts are given. 
In the brain extraction task using MRI, \citet{mohapatra2023brain} compared SAM with the brain extraction tool (BET) of the FMRIB Software Library. 
Quantitative results showed that the segmentation results of SAM were better than those of BET, demonstrating the potential of SAM for application in brain extraction tasks. 
\citet{deng2023segment} assessed the performance of SAM in digital pathology segmentation tasks, including tumor, non-tumor tissue and cell nuclei segmentation on whole-slide imaging. 
The results suggested that SAM delivers outstanding segmentation results for large connected objects. Nevertheless, it may not consistently achieve satisfactory performance for dense instance object segmentation, even with prompts of all the target boxes or 20 points per image. 
\citet{zhou2023can} applied SAM to the polyp segmentation task using five benchmark datasets under the \textit{Everything} setting.
The results showed that although SAM can accurately segment the polyps in some cases, a large gap exists between SAM and the state-of-the-art methods.
Additionally, \citet{liu2023samm} equipped the 3D Slicer software~\citep{pieper20043d} with SAM to assist in the development, assessment, and utilization of SAM on medical images.

Most recently, several studies tested the SAM on $\ge$10 public MIS datasets or tasks~\citep{he2023accuracy,mazurowski2023segment, ma2023segment,wu2023medical}.
In~\citep{he2023accuracy}, it was concluded that SAM's zero-shot segmentation performance is considerably inferior to that of traditional deep learning-based methods.
In~\citet{mazurowski2023segment}, the authors evaluated SAM's performance using different numbers of point prompts.
They observed that as the number of points increases, the performance of SAM converges.
They also noticed that SAM's performance is 1) overall moderate and 2) extremely unstable across different datasets and cases. 
\citet{ma2023segment} validated that the original SAM may fail on lots of medical datasets with the mean DICE score of 58.52\%. 
They then finetuned the SAM using medical images and found that the proposed MedSAM achieved a 22.51\% improvement on DICE compared with the SAM.
\citet{wu2023medical} adopted the \textit{Adapter} technique to finetune the SAM and enhance its medical ability.
Experiments validated that their proposed Medical SAM Adapter can outperform the state-of-the-art (SOTA) MIS methods (\textit{e.g.,} nnUnet~\citep{isensee2021nnu}). 

Although the above works investigated the performance of SAM in MIS, they had at least one of the following limitations:
\begin{enumerate}
	\item Small datasets. 
	Previous studies have only evaluated SAM’s performance in modalities, such as MRI, CT, and digital pathology.
	They included a limited number of segmented objects. 
	However, medical images contain multiple modalities and numerous anatomical structures or other objects requiring segmentation. 
	This has limited the comprehensive analysis in the field of MIS of the above studies~\citep{ji2023sam, ji2023segment, mohapatra2023brain, deng2023segment,zhou2023can}.
	\item Single SAM testing strategy. 
	Most previous studies~\citep{ji2023sam,mazurowski2023segment,zhou2023can} evaluated SAM with limited or even only one type of testing mode/strategy.
	However, different medical objects often exhibit different characteristics and thus may have their own suitable modes for testing. 
	The limited testing strategy may lead to inaccurate and incomplete analysis for SAM.
	\item Lack of comprehensive and in-depth assessments. 
	Some of the existing works~\citep{ji2023segment} only assessed the SAM via the visualization results provided by the online demo\footnote{\url{https://segment-anything.com/demo}}.
	Additionally, some studies focused only on limited metrics (\textit{e.g.,} DICE or IOU) to evaluate the performance of SAM~\citep{deng2023segment}.
	Most of the studies did not investigate SAM's perception of medical objects.
	Thus, the correlation between SAM's segmentation performance and medical objects' attributes has not been conducted carefully~\citep{he2023accuracy,mazurowski2023segment, ma2023segment}.
\end{enumerate}

The analysis of medical object perception is crucial. It can help the community better understand the factors that influence SAM segmentation performance (\textit{i.e.,} the ability to perceive medical objects), to better develop a new generation of general medical segmentation models.

In this report, we build a large medical image dataset named \textit{\ourdata}, including $1050K$ images with 18 different modalities (see Fig.~\ref{fig:modal_more_data}) and 84 objects (\textit{e.g.,} anatomical structures, lesions, cells, tools, etc.), to cover the entire body (see Fig.~\ref{fig:organ_data}).
This can help us comprehensively analyze and evaluate SAM's performance on medical images.
We then fully explore the different testing strategies of SAM and provide rich quantitative and qualitative experimental results to show SAM's perception of medical objects.
Finally, we deeply evaluate the correlation between SAM's performance and the characteristics (\textit{e.g.,} complexity, contrast, and size) of the objects.
We hope that this comprehensive report can provide the community with some insights into the future development of medical SAM.

\begin{figure*}[!ht]
	\centering
	\includegraphics[width=0.9 \linewidth]{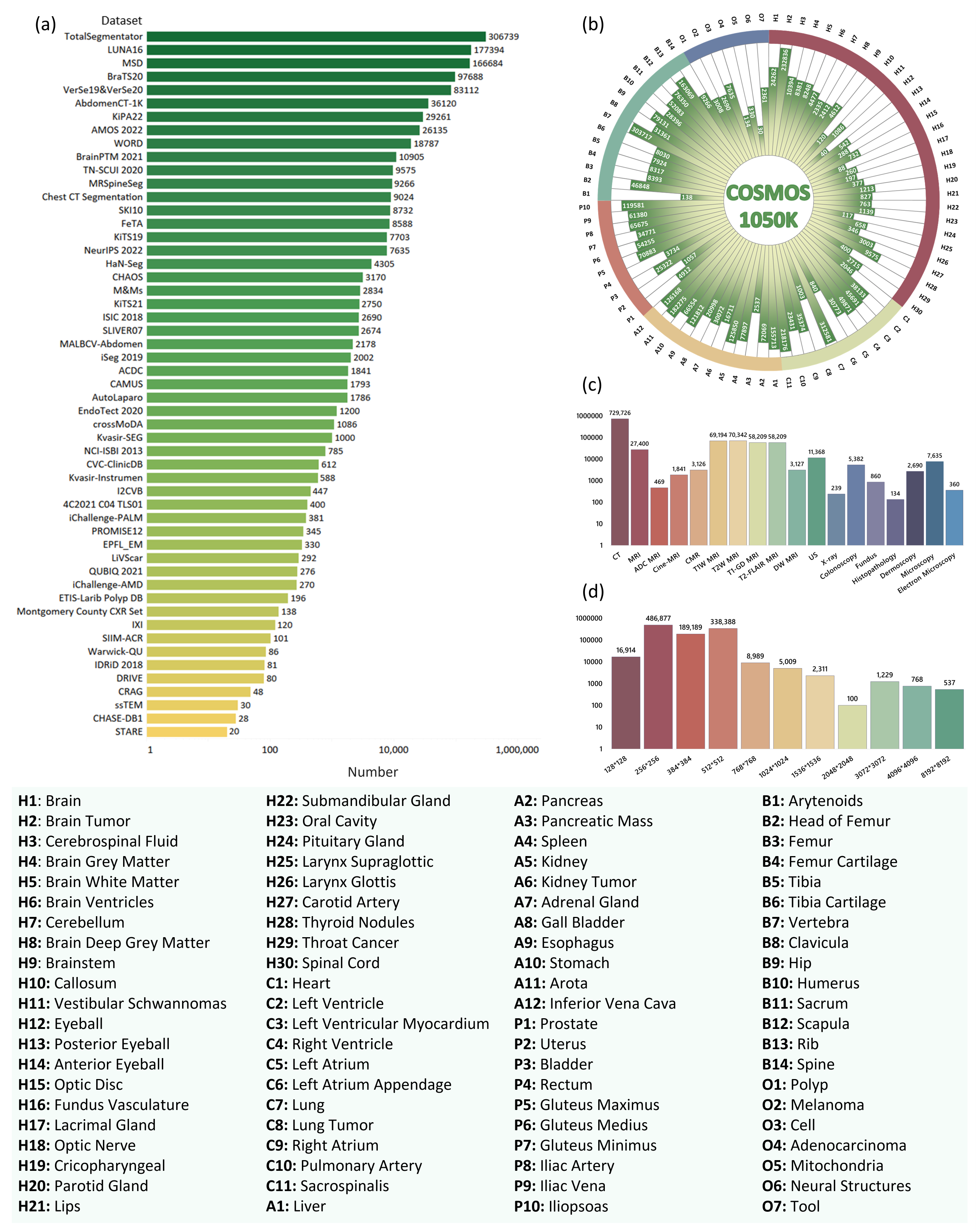}
        \caption{Statistics of \textit{\ourdata}~dataset. (a) Number of datasets after preprocessing. (b) Histogram distribution of 84 objects' quantity, as indicated by the abbreviated mapping provided in the legend. (c) Number of Modalities. (d) Histogram distribution of image resolutions. In (d), each bar represents an area interval distribution, \textit{e.g.,} $128*128$ represents the image area interval (0, $128*128$); $256*256$ represents the image area interval ($128*128$, $256*256$).}
	\label{fig:dataset}
\end{figure*}

\section{Dataset}
Medical images have various modalities such as CT, MRI, ultrasound (US), and X-ray. 
There are large domain gaps between different modalities~\citep{wang2020deep}, and various modalities have their advantages in visualizing specific objects, including anatomical structures and lesions~\citep{liu2023samm}. 
To fully evaluate the generalization performance of SAM in MIS, we collected 53 public datasets and standardized them to construct the large \textit{\ourdata}~dataset.
For the categorization system (\textit{e.g.,} modal categorization) of \textit{\ourdata}, we referred to the official introduction of each public dataset and the recently-published study~\citep{butoi2023universeg} (see Table~\ref{tab:datasetdescription} for more details).
Fig.~\ref{fig:modal_more_data} and Fig.~\ref{fig:organ_data} illustrate various imaging modalities and most of the clinical segmentation objects covered in the dataset, respectively.
We describe the details of \textit{\ourdata}~in the following two aspects, including image collection and preprocessing specification.\par

\begin{table*}[!ht]
\tiny
    \resizebox{\textwidth}{!}{
	\begin{threeparttable}[b]
		\centering
		\scriptsize\caption{Description of the collected dataset. MALBCV-Abdomen is an abbreviation for the abdomen dataset of Multi-Atlas Labeling Beyond the Cranial Vault.}
		
			\begin{tabular}{lll}
				\toprule
				Dataset Name & Description &  Image Modalities \\
				\midrule    
				AbdomenCT-1K~\citep{ma2021abdomenct} & Liver, kidney, spleen and pancreas & CT \\
				ACDC~\citep{bernard2018deep} & Left and right ventricle and left ventricular myocardium & Cine-MRI\\
				AMOS 2022~\citep{ji2022amos} & Abdominal multi-organ segmentation & CT, MRI \\
				AutoLaparo~\citep{wang2022autolaparo} & Integrated dataset with multiple image-based perception tasks & Colonoscopy \\
				BrainPTM 2021~\citep{avital2019neural, nelkenbaum2020automatic} &  white matter tracts & T1W MRI, DW MRI \\
				BraTS20~\citep{menze2014multimodal, bakas2017advancing, bakas2018identifying} & Brain tumor & T1W MRI, T2W MRI, T1-GD MRI, T2-FLAIR MRI\\
				CAMUS~\citep{leclerc2019deep} & Four-chamber and Apical two-chamber heart & US \\
				CellSeg Challenge-NeurIPS 2022~\citep{NeurIPS22CellSeg} & Cell segmentation & Microscopy \\			
   CHAOS~\citep{kavur2021chaos} & Livers, kidneys and spleens & CT, T1W MRI, T2W MRI\\
				CHASE-DB1~\citep{zhang2016robust} & Retinal vessel segmentation & Fundus \\
				Chest CT Segmentation~\citep{chestsegmentation} & Lungs, heart and trachea & CT\\
				CRAG~\citep{graham2019mild} & Colorectal adenocarcinoma  & Histopathology \\
				crossMoDA~\citep{shapey2019artificial, shapey2021segmentation} & Vestibular schwannoma & T1W MRI\\
				CVC-ClinicDB~\citep{bernal2015wm} & Polyp & Colonoscopy \\
				DRIVE~\citep{liu2022full} & Retinal vessel segmentation & Fundus \\
				EndoTect 2020~\citep{hicks2021endotect} & Polyp & Colonoscopy \\
				EPFL-EM~\citep{lucchi2013learning} & Mitochondria and synapses segmentation & Electron Microscopy \\
				ETIS-Larib Polyp DB~\citep{bernal2017comparative, yoon2022colonoscopic} & Polyp & Colonoscopy \\
				FeTA~\citep{payette2021automatic} & Seven tissues of the infant brain & T2W MRI \\
				HaN-Seg~\citep{podobnik2023han} & Healthy organs-at-risk near the head and neck & CT \\
				I2CVB~\citep{lemaitre2015computer} & Prostate & T2W MRI\\
				iChallenge-AMD~\citep{li2020self} & Optic disc and fovea & Fundus \\
				iChallenge-PALM~\citep{huazhu2019palm} & Optic disc and lesions from  pathological myopia patients & Fundus \\
				IDRiD 2018~\citep{porwal2018indian} & Optic disc, fovea and lesion segmentation & Fundus \\
				iSeg 2019~\citep{sun2021multi} & White matter, gray matter, and cerebrospinal fluid of infant brain & T1W MRI, T2W MRI \\
				ISIC 2018~\citep{tschandl2018ham10000,codella2018skin,codella2019skin} & Melanoma of skin & Dermoscopy \\
				IXI~\citep{ixi} & Callosum & T1W MRI\\
				KiPA22~\citep{he2021meta,he2020dense,shao2011laparoscopic,shao2012precise} & Kidney, tumor, renal vein and renal artery & CT \\
				KiTS19~\citep{heller2020kits19} & Kidneys and tumors & CT\\
				KiTS21~\citep{zhao2022coarse} & Kidneys, cysts, tumors, ureters, arteries and veins & CT\\
				Kvasir-Instrumen~\citep{10.1007/978-3-030-67835-7_19} & Gastrointestinal procedure instruments such as snares, balloons, etc. & Colonoscopy \\
				Kvasir-SEG~\citep{jha2020kvasir} & Gastrointestinal polyp & Colonoscopy\\
				LiVScar~\citep{karim2016evaluation} & Infarct segmentation in the left ventricle & CMR \\
				LUNA16~\citep{setio2017validation} & Lungs, heart and trachea & CT \\
				M\&Ms~\citep{campello2021multi} & Left and right ventricle and left ventricular myocardium & CMR\\
				MALBCV-Abdomen~\citep{landman2015miccai} & Abdominal multi-organ segmentation & CT \\
				Montgomery County CXR Set~\citep{jaeger2014two} & Lung & X-ray\\
				MRSpineSeg~\citep{pang2020spineparsenet} & multi-class segmentation of vertebrae and intervertebral discs & MRI \\
				MSD~\citep{MSD,simpson2019large} & Large-scale collection of 10 Medical Segmentation Datasets & CT, MRI, ADC MRI, T1W MRI, T2W MRI, T1-GD MRI, T2-FLAIR MRI\\
				NCI-ISBI 2013~\citep{li2013automated} & Prostate (peripheral zone, central gland) & T2W MRI\\
                PROMISE12~\citep{litjens2014evaluation} & Prostate & T2W MRI\\
				QUBIQ 2021~\citep{ji2021learning} & Kidney, prostate, brain growth, and brain tumor & CT, MRI, T1W MRI, T2W MRI, T1-GD MRI, T1-FLAIR MRI\\
				SIIM-ACR~\citep{SIIM-ACR,viniavskyi2020weakly} & Pneumothorax segmentation & X-ray \\
				SKI10~\citep{lee2010learning} & Cartilage and bone segmentation from knee data & MRI \\
				SLIVER07~\citep{heimann2009comparison} & Liver  & CT \\
				ssTEM~\citep{cardona2010integrated} & Neuronal structures & Electron Microscopy \\
				STARE~\citep{hoover2000locating, hoover2003locating} & Retinal vessel segmentation & Fundus \\
				TN-SCUI 2020~\citep{jianqiao_zhou_2020_3715942} & Thyroid nodule & US \\
				TotalSegmentator~\citep{wasserthal2022totalsegmentator} &  Multiple anatomic structures segmentation(27 organs, 59 bones, 10 muscles, and 8 vessels) & CT \\
                VerSe19~\citep{sekuboyina2021verse} & Spine or vertebral segmentation & CT \\ VerSe20~\citep{loffler2020vertebral, liebl2021computed} & Spine or vertebral segmentation & CT \\
				Warwick-QU~\citep{sirinukunwattana2017gland} & Gland segmentation & Histopathology \\
				WORD~\citep{luo2022word} & Abdominal multi-organ segmentation & CT \\   
				4C2021 C04 TLS01~\citep{tls} & Throat and hypopharynx cancer lesion area & CT\\
				\bottomrule
		\end{tabular}
        \label{tab:datasetdescription}%
	\end{threeparttable}}
 \vspace{5mm}
\end{table*}%

\begin{figure}[!ht]
	\centering
	\includegraphics[width=1 \linewidth]{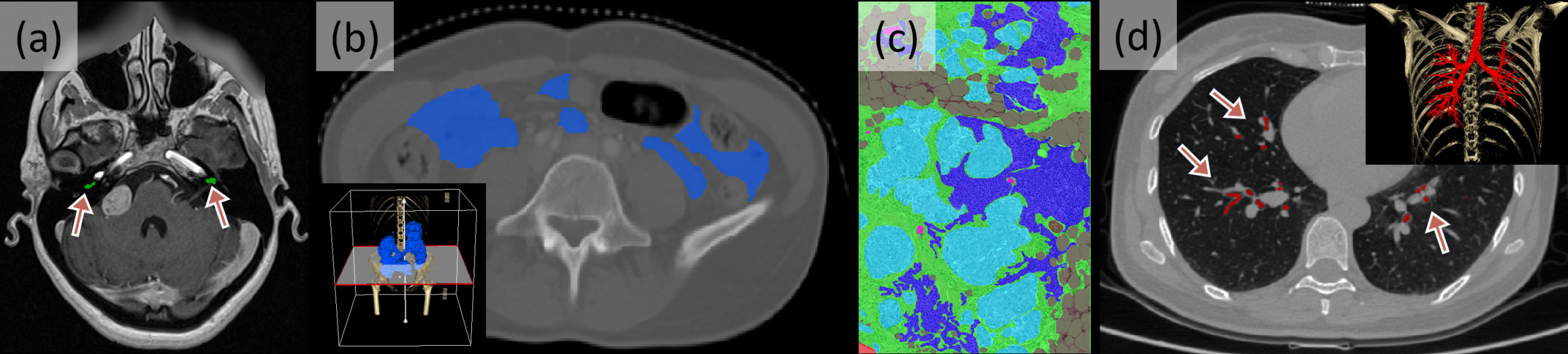}
	\caption{Typical examples of meeting the exclusion criteria. (a) cochlea (criteria 1), (b) intestine (criteria 2), (c) histopathological breast cancer (criteria 3), and (d) lung trachea trees (criteria 3). The corners (b) and (d) show the 3D rendering images obtained by \textit{Pair} annotation software package~\citep{liang2022sketch}.}
	\label{fig:collect}
\end{figure}

\begin{figure*}[!ht]
	\centering
	\includegraphics[width=1 \linewidth]{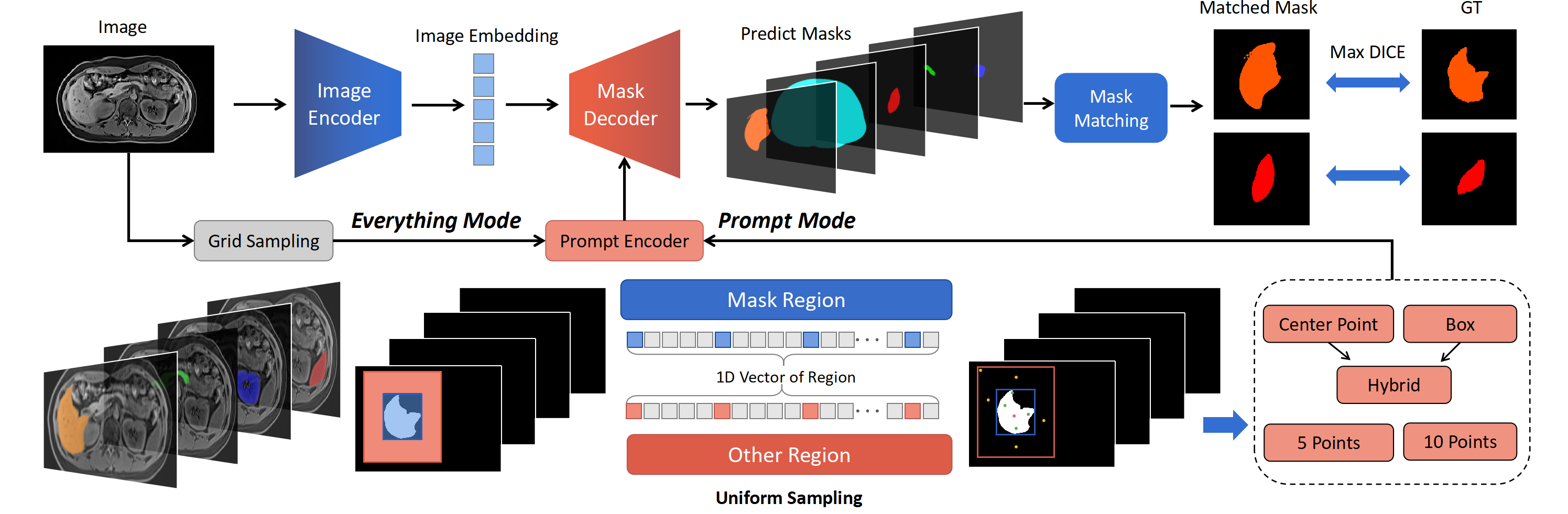}
	\caption{Testing pipeline of SAM in our study.}
	\label{fig:framework}
\end{figure*}

\subsection{Dataset Collection}
Medical images cover a wide range of object types, such as brain organs and tumors~\citep{simpson2019large, shapey2019artificial,bakas2018identifying,sun2021multi,podobnik2023han}, lungs and hearts~\citep{bernard2018deep,jaeger2014two,viniavskyi2020weakly,setio2017validation}, abdomen~\citep{simpson2019large,zhao2022coarse,ma2021abdomenct,ji2022amos,luo2022word}, spine~\citep{sekuboyina2021verse,loffler2020vertebral,pang2020spineparsenet}, cells~\citep{lee2022mediar,NeurIPS22CellSeg}, and polyps~\citep{jha2020kvasir,hicks2021endotect}. Table~\ref{tab:datasetdescription} presents a detailed list of the collected MIS datasets and Fig.~\ref{fig:dataset} (a) shows the amount of each dataset after preprocessing.
To be compatible with different modes of evaluating SAM, we employed the following exclusion criteria: 
1) Exclude objects that are extremely small, such as the cochlea, shown in Fig.~\ref{fig:collect} (a), and ureter. 
This is due to the difficulty of automatically generating points or box prompts on extremely small objects. 
2) Exclude objects in the 3D volume where their overall target became significantly separated as the slice was sequentially extracted, such as the intestine (seen in Fig.~\ref{fig:collect} (b)), mandible, and thyroid gland. We aim to avoid confusing the main object and generate unique boxes for each object. 
3) Exclude objects with a relatively discrete overall structure, such as histopathological images of breast cancer (see Fig.~\ref{fig:collect} (c)), slices of lung trachea trees (see Fig.~\ref{fig:collect} (d)), renal arteries, and veins. Most of these objects are dispersed into multiple items in a 2D slice and embedded in other objects, resulting in a failure to sensibly employ SAM's prompt mode on those objects to verify. 
According to the above criteria, \textit{\ourdata}~now comprises a total of 84 objects, with their numbers depicted in Fig.~\ref{fig:dataset} (b). These objects are categorized only once in one image, without distinguishing locations or detailed divisions (\textit{e.g.}, 'Left Lung' and 'Right Lung' are grouped as 'Lung', and various instruments are considered as 'Tool'). 
Further details can be found in the legend of Fig.~\ref{fig:dataset}. 
The histogram distributions of modalities and image resolutions are shown in Fig.~\ref{fig:dataset} (c) and Fig.~\ref{fig:dataset} (d), respectively. Given the significant variation of the same object across different modalities, including differences in gray distribution and texture features, we further divide them into 125 object-modality paired targets.\par

\subsection{Dataset Preprocessing Specification}
\ourdata~contain different labels, modalities, formats, and shapes from one to another. Additionally, the original version of SAM only supported 2D input, and 2D format is fundamental and even a basic component for 3D/4D format.
To standardize the data across different datasets, we applied the following preprocessing steps for each collected public dataset. 

For 3D volumes, the whole procedure can be summarized as follows: 
1) Extract slices along the main viewing plane owing to its higher resolution.
In CT, it is usually a transverse plane, while in MRI, it may be a transverse plane, \textit{e.g.,} prostate, brain tumor, or a sagittal plane, \textit{e.g.,} spine and heart.
2) Retain slices with the sum of the pixel values of their labels greater than 50 for any 3D image and label volumes. This ensures that each slice has the corresponding correct label.
3) Normalize the extracted image intensities by \textit{min-max} normalization: $I_n = 255*(I-I_{min})/(I_{max}-I_{min})$, limiting the range to (0, 255). $I$ means the original extracted image, and $I_n$ represents the normalized image. $I_{min}$ and $I_{max}$ are the minimum and maximum intensity values of $I$. Simultaneously, we reset the pixel values of the mask according to the object's category or location (\textit{e.g.,} left and right kidney have different pixel values).
This is due to the fact that medical images may vary widely in the voxel or pixel values.  
Examples include MRI with an intensity range of (0, 800) and CT with an intensity range of (-2000, 2000), while other modalities may already be in the range of (0, 255)~\citep{butoi2023universeg}.
4) Save images and labels in PNG format. 
For 4D data (N, W, H, D), we convert the data into N sets of 3D volume and then follow the 3D volume processing flow. Here, N represents the number of paired volumes within the 4D data. 

For 2D images, the preprocessing is as follows: 
1) Retain images with the sum of the pixel values of their labels greater than 50.
2) Reset the pixel value of labels according to the object category or location within the range of 1 to 255. 
For the CellSeg Challenge-NeurIPS 2022~\citep{NeurIPS22CellSeg}, owing to the wide range of original label values (1-1600), we reconstruct each image and label into several sub-figures to ensure a uniform label range.
3) Convert the format of images and labels from BMP, JPG, TIF, etc. to PNG for achieving consistent data loading.\par
In total, \ourdata~consists of 1,050,311 2D images or slices, with 1,003,809 slices originating from 8,653 3D volumes and 46,502 being standalone 2D images. Furthermore, the dataset incorporates 6,033,198 masks.

\section{Methodology}
\subsection{Brief Introduction to SAM}
SAM diverges from traditional segmentation frameworks by introducing a novel promptable segmentation \textit{task}, which is supported by a flexible prompting-enabled \textit{model} architecture and vast and diverse sources of training \textit{data}.
A \textit{data engine} was proposed to build a cyclical process that utilizes the model to facilitate data collection and subsequently leverages the newly collected data to enhance the model's performance.
Finally, SAM was trained on a massive dataset comprising over one billion masks from 11 million licensed 2D images.

As shown in Fig.~\ref{fig:framework}, SAM mainly contains three components: an image encoder, a prompt encoder, and a mask decoder.
The image encoder was, with the backbone of ViT, pre-trained by the masked autoencoder (MAE~\citep{he2022masked}) technique.
It takes one image as input and outputs the image embedding for combinations with subsequent prompt encoding.
The prompt encoder consists of dense (masks) and sparse (points, boxes, and text) branches.
The dense branch encodes the mask prompts via a convolutional neural network (CNN).
For the sparse one, the points and boxes can be represented by positional encoding~\citep{tancik2020fourier}, while the text is embedded by CLIP~\citep{radford2021learning}.
Finally, the mask decoder decodes all the embeddings and predicts the masks.


During testing, SAM supports both automatic \textit{Everything} and manual \textit{Prompt} modes.
For the former, the users only have to input an image to SAM, and then all the predicted masks will be produced automatically.
For the latter, the users manually provide some additional hints to SAM, including masks, boxes, points, and texts, to give SAM more information about the segmented objects.
Details of these two modes are presented in the next sub-sections.
It is noted that SAM can only find multiple targets in the image, without outputting their detailed categories (i.e., single-label: object or not).

In the official GitHub repository\footnote{\url{https://github.com/facebookresearch/segment-anything}\label{fn:github}}, authors provide three types of pre-trained models with different backbone sizes, named \textit{ViT-B}, \textit{ViT-L}, and \textit{ViT-H}.
Their model parameters range from small to large.
In~\citet{kirillov2023segment}, \textit{ViT-H} shows substantial performance improvements over \textit{ViT-B}. However, the former requires multiplied testing time due to its increased complexity.

While evaluating SAM in medical images, one study used six medical datasets and found that there is no obvious winner among \textit{ViT-B}, \textit{ViT-L}, and \textit{ViT-H}~\citep{mattjie2023zeroshot}.
In our study, we chose the smallest \textit{ViT-B} (with 12 transformer layers and 91M parameters) and largest \textit{ViT-H} (with 32 transformer layers and 636M parameters) as encoders to run all testing modes.
We hope that a comprehensive evaluation of models of different sizes on large \textit{\ourdata} can provide more inspirations for researchers.

\subsection{Automatic Everything Mode} 
In the \textit{Everything} mode (\textbf{$S_1$}), SAM produces segmentation masks for all the potential \textit{objects} in the whole image without any manual priors.
The initial step of the process involves generating a grid of point prompts (\textit{i.e.,} grid sampling) that covers the entire image. 
Based on the uniformly sampled grid points, the prompt encoder will generate the point embedding and combine it with the image embedding. 
Then, the mask decoder will take the combination as input and output several potential masks for the whole image.
Subsequently, a filtering mechanism is applied to remove duplicate and low-quality masks using the techniques of confidence score, stability evaluation based on threshold jitter, and non-maximal suppression (NMS).

\subsection{Manual Prompt Mode}
In the \textit{prompt} mode, SAM provides different types of prompts, including points, boxes, and texts.
The point prompt covers both positive and negative points, which indicate the foreground and background of one object, respectively.
The box prompt represents the spatial region of the object that needs to be segmented.
Furthermore, the text prompt indicates one sentence (i.e., basic information in terms of position, color, size, etc.) to describe the object.
Notably, the text prompt has not been released yet on the official GitHub repository$^{\ref{fn:github}}$. 

As shown in Fig.~\ref{fig:framework}, our prompt mode contains five strategies, including \textbf{\textit{one positive point}} (\bm{$S_2$}), \textbf{\textit{five positive points}} (\bm{$S_3$}), \textbf{\textit{five positive points with five negative points}} (\bm{$S_4$}), \textbf{\textit{one box}} (\bm{$S_5$}), and \textbf{\textit{one box with one positive point}} (\bm{$S_6$}).
We further established a unified rule for point selection to ensure randomness, repeatability, and accuracy.
For the positive point selection, 
a) we first calculated the center of mass of the ground truth (GT) mask (red point in Fig.~\ref{fig:framework}).
b) If the center of mass was inside the GT mask, we took the center as the first positive point.
c) Then, we directly flattened the GT mask to a one-dimensional vector and obtained the other positive points by adopting the uniform sampling method (green points in Fig.~\ref{fig:framework}).
d) If the center of mass was outside the GT mask, all the required positive points would be obtained by performing step c.
For the negative point selection, we aimed to avoid selecting points that were too distant from the target region.
Specifically, we first enlarged the bounding box of the GT by two times.
The negative points were generated by even sampling in the non-GT region (yellow points in Fig.~\ref{fig:framework}).
Last, for the box selection, we directly adopted the bounding box of the GT mask without any additional operations.
The above strategy can ensure the repeatability of the experiments. Besides, we tend to test the theoretical optimum performance of SAM via the selection of the center of mass and tight box. Since they may include the most representative features of the target. It is noted that SAM allows multiple prompts to be input into the network once. Thus, for a fair comparison, we test the one-round interaction performance of SAM under the above five prompt strategies ($S_2$-$S_6$).

\subsection{Inference Efficiency}
We performed multiple tests ($n$) on an image using different strategies to obtain a final assessment (see Fig.~\ref{fig:framework}).
In SAM’s original code logic and design, the same encoding operation is required on one image $n$ times, which results in poor runtime efficiency in our multi-strategy test scenario.
The situation becomes even worse when using high-resolution inputs. 
Based on this observation, we computed the embedding features of all input images in advance and saved them as intermediate files.
Accordingly, the image embeddings could be reused to relieve the computational burden of the inference pipeline.
Thereby, the overall efficiency of the SAM testing could be improved by nearly \textit{n} times.
The more testing strategies in SAM, the more time that could be saved.
This can simply be extended to other multi-strategy testing scenarios of SAM.
\subsection{Mask-matching Mechanism for Segmentation Evaluation}
SAM generated multiple binary masks for each input image, but not all of them included the corresponding object. Hence, we proposed a mask-matching mechanism to evaluate segmentation performance using SAM in each mode. 
Specifically, for an object (one of the foregrounds) in a given image, we calculated a set of dice scores \{DICE$_n$\}$_{n=1}^N$ between \textit{N} binary predicted masks \{\textbf{P}$_n$\}$_{n=1}^N$ and the GT \textbf{G}. 
Then, the one with the highest dice score in the set was selected as the matched predicted mask \textbf{P} for subsequent segmentation evaluation.
This process for obtaining \textbf{P} can be expressed as follows:
\begin{equation}
	\textbf{P} = max\{(\textbf{P}_1 \cdot \textbf{G}), (\textbf{P}_2 \cdot \textbf{G}), \ldots, (\textbf{P}_N \cdot \textbf{G})\},
\end{equation}
where \textit{N} is the total number of predicted binary masks for an object in one image. The operations $(\cdot)$ and $max\{\}$ indicate computing a dice score between one predicted mask and the GT, while $max$ denotes obtaining the predicted mask with the highest dice score.

\section{Experiments and Results}

\subsection{Implementation Details}
\textbf{Code Implementation and Logits.}
In this study, we implemented the testing pipeline of SAM basically following the official GitHub repository$^{\ref{fn:github}}$.
For our multi-strategy testing scenario, we ran the SAM algorithm \textit{n} times and extracted image embeddings \textit{n} times. We observed that the process of image embedding extraction was time-consuming. However, since the same embedding could be reused for different testing strategies, we sought to optimize and accelerate this multi-extraction process. Thus, we refactored part of the code. For each test image, we only used the image encoder for feature extraction once and saved the embedded features as an \textit{npy} file.
When different testing strategies were applied, only the corresponding \textit{npy} files needed to be loaded, which significantly improved the testing efficiency (approximately $\textit{n}\times$).
Additionally, for the prompt testing, we calculated the required points and boxes once after image embedding and stored them as \textit{npz} files.
Thereby, all the prompt testing strategies could directly use the \textit{npz} information without recalculation.

\textbf{Package Versions and Features.}
We used multiple GPUs, including the NVIDIA GTX 2080Ti with 12GB, NVIDIA GTX 3090 with 24GB, and NVIDIA A40 with 48GB, for testing.
We implemented the SAM with \textit{python}~(version 3.8.0), \textit{PyTorch}~(version 2.0.0) and \textit{torchvision}~(version 0.15.1). We further adopted: 1) \textit{torch.compile} with the \textit{max-autotune} mode to pack the model; 2) \textit{torch.cuda.amp} to adaptively adjust the types of tensors to Float16 or Float32; 3) \textit{@torch.inference\_mode} to replace the common \textit{torch.no\_grad} to reduce GPU memory occupation and improve inference speed while keeping model calculation precision.
We used \textit{Numpy} package (version 1.24.2) to generate/calculate the prompts (boxes/points) for simulating the human interaction process (click, draw box, etc.) with the testing images. Besides, we used \textit{OpenCV} (version 4.7.0.72) and \textit{Matplotlib} (version 3.7.1) software packages for visualizing provided prompts and final results.

\textbf{Testing Strategy Design.}
We designed different settings to fully explore the performance of SAM under various testing strategies.
We first considered the \textit{\textbf{Everything}} mode, which is the key feature of SAM.
It can output all the predicted masks for a given image without any manual hints.
Considering that MIS is an extremely challenging task, we progressively incorporated various manual prompts to aid in accurate segmentation.
The manual prompts included 1) one positive point, 2) five positive points, 3) five positive and five negative points, 4) one box, and 5) one box and one positive point.
The manually provided prompts can better guide the SAM to output accurate segmentation results.
We summarize all the testing strategies and their abbreviations as:

\begin{itemize} 
\setlength{\itemsep}{0.1pt}
\item \textbf{$S_1$: automatic \textit{everything} mode;}
\item \textbf{$S_2$: one positive point;}
\item \textbf{$S_3$: five positive points;}
\item \textbf{$S_4$: five positive and five negative points;}
\item \textbf{$S_5$: one box;}
\item \textbf{$S_6$: one box and one positive point.}
\end{itemize}

\subsection{Evaluation Metrics}
To fully evaluate SAM's segmentation performance, we used three common metrics, as shown below:

\begin{enumerate}
	\item \textbf{DICE Coefficient (DICE, \%)}~\citep{dice1945measures,bilic2023liver}: A similarity measure to evaluate the overlap between the prediction and GT. Ranging from [0, 1], a higher value indicated a better performance of the model.
	\item \textbf{Jaccard Similarity Coefficient (JAC, \%)}~\citep{jaccard1908nouvelles,qian2022hasa}: Also called IOU, is used to measure the similarity between two masks.
          It is similar to DICE, but with a different calculation method. 
          Specifically, for the predicted and GT masks, A and B, JAC calculates the intersection ($|A\cap B|$) over union ($|A\cup B|$).
          JAC ranges from 0 to 1, and higher values indicate better performance.
	\item \textbf{Hausdorff Distance (HD, \textit{pixel})}~\citep{crum2006generalized,karimi2020reducing}: A measure that evaluates the degree of similarity between two sets of points, which can reflect the distance between each point in the prediction to the points in the GT. It is more sensitive to the boundary than DICE.

\end{enumerate}

In the following experiments, we will primarily analyze the performance in terms of DICE and HD.
The results of another evaluation index for similarity measurement (i.e., JAC) are shown in the supplementary materials.

\begin{figure}[!t]
	\centering
	\includegraphics[width=1 \linewidth]{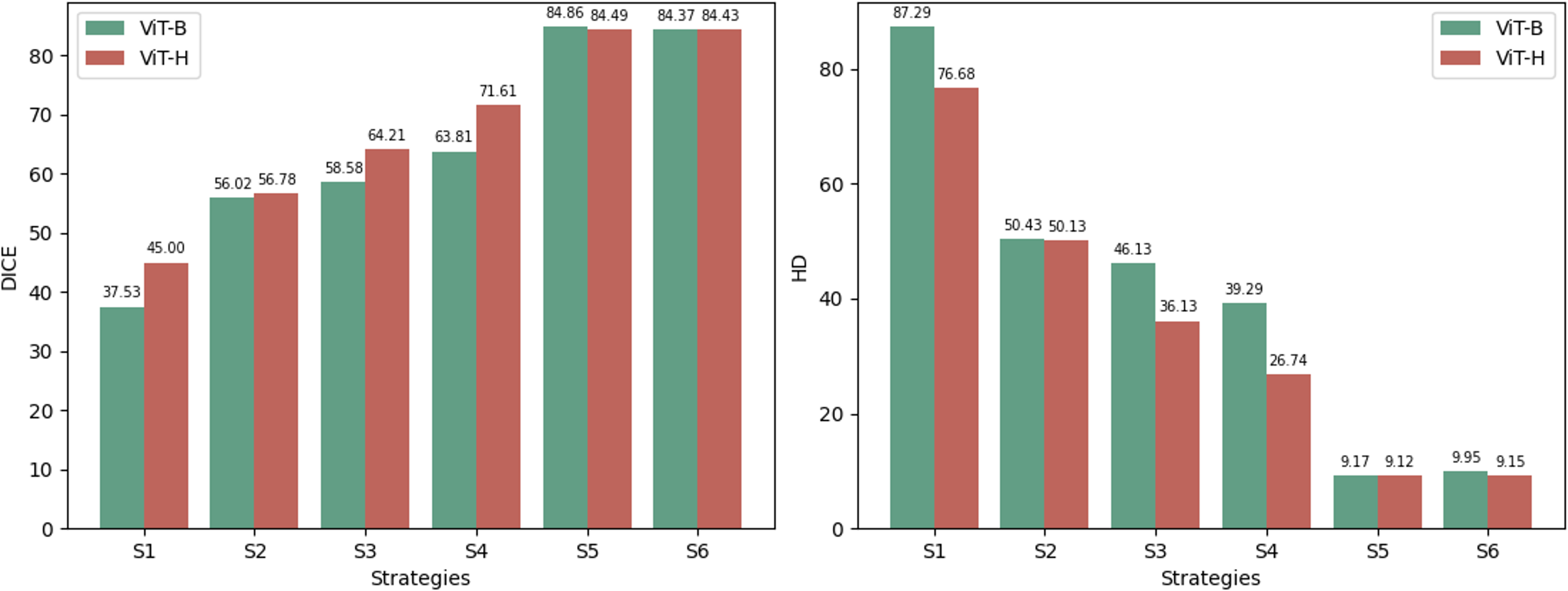}
	\caption{Comparison of the average performance of ViT-B and ViT-H under different strategies.}
	\label{fig:strategy_dice_hd}
\end{figure}

\begin{figure*}[!t]
	\centering
	\includegraphics[width=1 \linewidth]{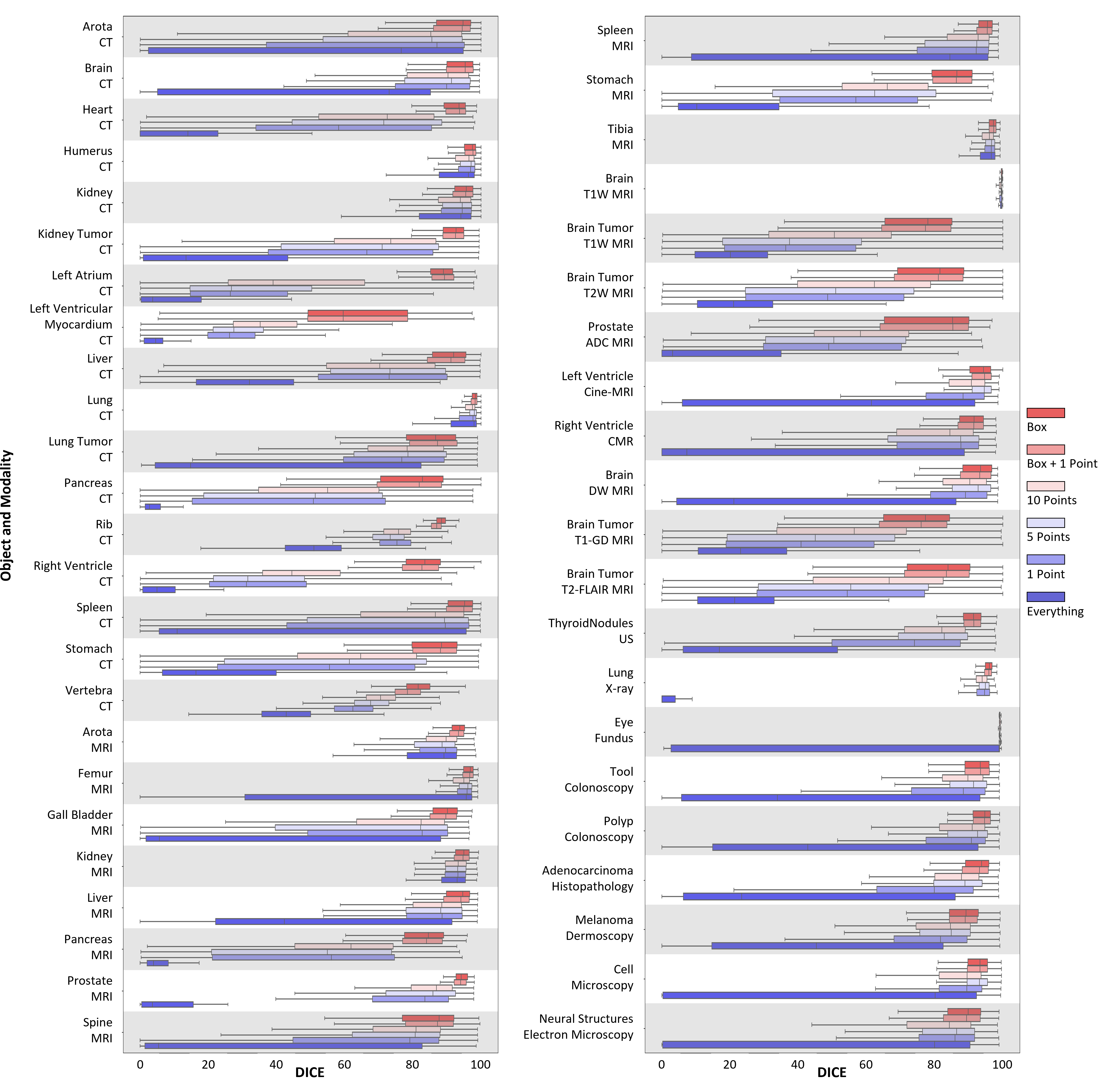}
	\caption{Comparison of DICE performance for selective common medical objects under ViT-B in different testing strategies.}
	\label{fig:choose_b}
\end{figure*}

\begin{figure*}[!t]
	\centering
	\includegraphics[width=1 \linewidth]{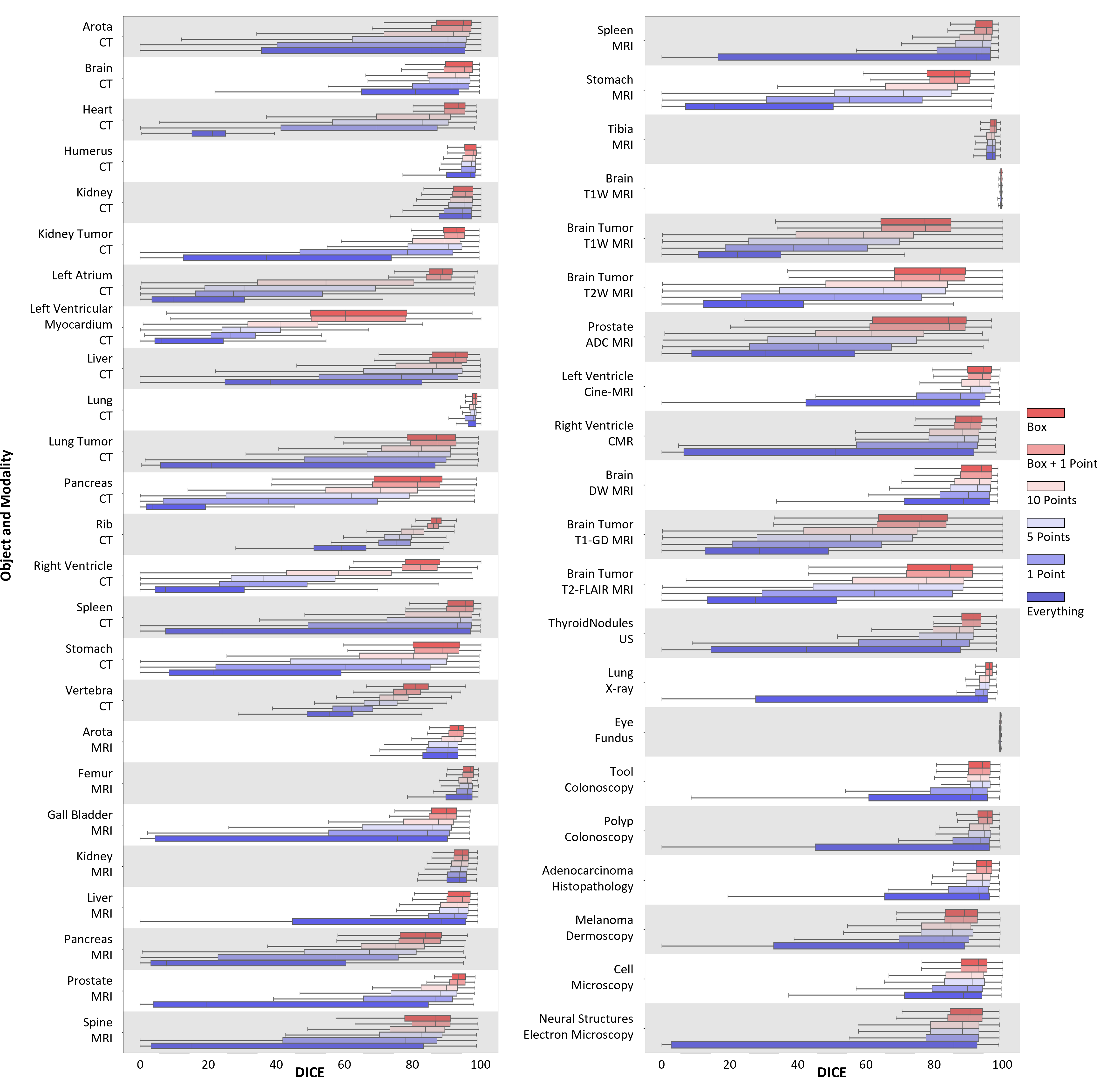}
	\caption{Comparison of DICE performance for selective common medical objects under ViT-H in different testing strategies.}
	\label{fig:choose_h}
\end{figure*}

\subsection{Segmentation Performance under Different Models}

\begin{figure*}[!h]
	\centering
	\includegraphics[width=0.8 \linewidth]{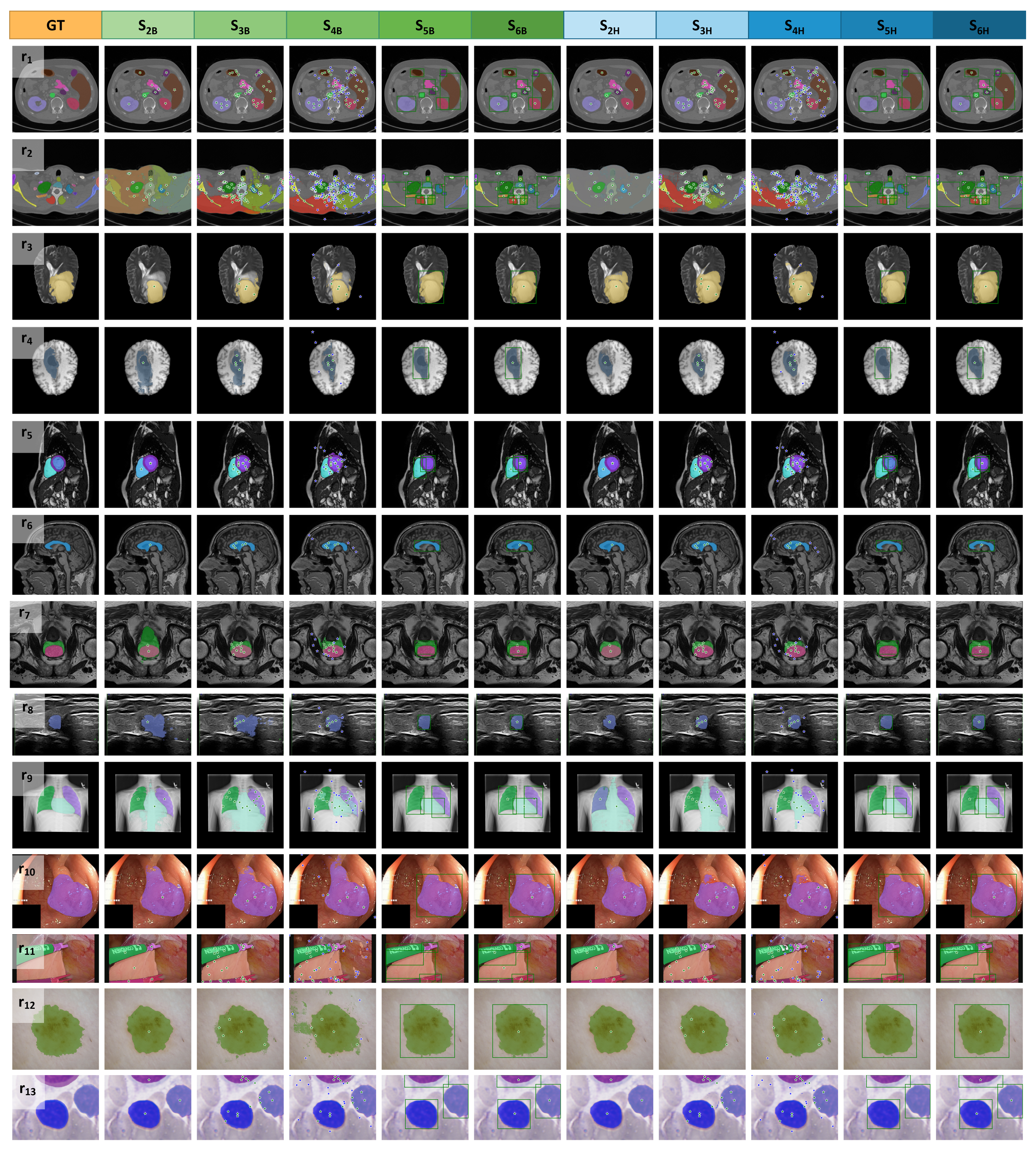}
	\caption{Typical good cases of SAM ($r$: row). $r_1$, $r_2$: CT, $r_3$, $r_7$: T2W MRI, $r_4$, $r_6$: T1W MRI, $r_5$: CMR, $r_8$: US, $r_9$: X-ray, $r_{10}$, $r_{11}$: Colonoscopy, $r_{12}$: Dermoscopy, $r_{13}$: Microscopy. Green and blue stars represent positive and negative point prompts, respectively. The green box indicates the box prompt.}
	\label{fig:good_data}
\end{figure*}

In this section, we tend to compare the segmentation performance between two models (\textit{ViT-B} and \textit{ViT-H}) under different strategies.
It can be observed from Fig.~\ref{fig:strategy_dice_hd} that, under \textit{Everything} mode ($S_1$), \textit{ViT-H} exceeds \textit{ViT-B} with 7.47\% at DICE and is 10.61 pixels lower than \textit{ViT-B} on HD.
For the \textit{one point} prompt ($S_2$), \textit{ViT-H} achieves slightly higher average performance than \textit{ViT-B}.
As the number of point prompts increases, the advantages of \textit{ViT-H} will become more obvious.
While for the resting strategies (box without/with one point, $S_5$-$S_6$), their performance are very close (differences in DICE: 0.37\% and 0.06\%).
Compared to the point prompt, the box prompt contains more region information about the object.
Thus, it can better guide SAM with different models to achieve better segmentation performance.
DICE and HD performance for specific objects can be found in Table~\ref{tab:dice_select} and Table~\ref{tab:hd_select}.
We only present the detailed performance of 40 objects in the main text, and full results can be found in supplementary materials.
Fig.~\ref{fig:choose_b}
and Fig.~\ref{fig:choose_h} display the DICE distribution for the same objects evaluated under varying model sizes. 
It proves that \textit{ViT-H} exhibits more stable performance with smaller standard deviations compared to \textit{ViT-B} (see Kidney-CT, Prostate-MRI, and Polyp-Colonoscopy for typical examples). 
See Fig.~\ref{fig:good_data} for prediction visualization.
Additional comparison and visualization results can be found in the supplementary materials.

\begin{table*}[!t]
  \centering
  \scriptsize
  \caption{Performance on selective common medical objects with different modalities in terms of DICE score (\%). \textit{ViT-B} and \textit{ViT-H} represent the small and large encoders of SAM. $S_1$-$S_6$ represent different testing strategies, including everything, 1 point, 5 points, 10 points, box, and box with 1 point, respectively.}
  \resizebox{\textwidth}{!}{
    \begin{tabular}{lcccccc|cccccc}
    \hline
       \multirow{2}[0]{*}{Object-Modality} & \multicolumn{6}{c|}{ViT-B}                    & \multicolumn{6}{c}{ViT-H} \\
    \cline{2-13}          
    & $S_{1}$   & $S_{2}$   & $S_{3}$   & $S_{4}$   & $S_{5}$   & $S_{6}$   & $S_{1}$   & $S_{2}$   & $S_{3}$   & $S_{4}$   & $S_{5}$   & $S_{6}$ \\
    \cline{0-12}      
   Arota-CT	&	57.87 	&	68.71 	&	71.92 	&	74.88 	&	86.88 	&	85.57 	&	65.71 	&	70.59 	&	76.68 	&	81.11 	&	86.80 	&	85.09 	\\
Brain-CT	&	54.79 	&	80.71 	&	82.11 	&	83.49 	&	91.34 	&	91.37 	&	72.70 	&	83.03 	&	85.59 	&	86.45 	&	91.14 	&	90.91 	\\
Heart-CT	&	13.01 	&	57.23 	&	63.53 	&	66.65 	&	90.15 	&	90.50 	&	22.33 	&	62.10 	&	70.63 	&	76.75 	&	90.06 	&	90.06 	\\
Humerus-CT	&	83.54 	&	91.66 	&	91.93 	&	91.91 	&	95.16 	&	95.07 	&	86.61 	&	91.93 	&	92.46 	&	93.31 	&	95.28 	&	95.17 	\\
Kidney-CT	&	75.27 	&	86.86 	&	87.28 	&	87.84 	&	93.60 	&	93.26 	&	82.88 	&	87.66 	&	89.49 	&	90.88 	&	93.29 	&	93.09 	\\
Kidney Tumor-CT	&	26.83 	&	60.57 	&	62.98 	&	68.93 	&	90.69 	&	90.63 	&	43.22 	&	67.30 	&	80.27 	&	83.23 	&	90.74 	&	90.95 	\\
Left Atrium-CT	&	13.90 	&	34.60 	&	35.46 	&	44.93 	&	87.50 	&	87.82 	&	22.84 	&	36.94 	&	40.74 	&	55.71 	&	87.23 	&	86.76 	\\
Left Ventricular Myocardium-CT	&	7.79 	&	27.59 	&	29.33 	&	37.12 	&	63.21 	&	63.16 	&	13.43 	&	27.97 	&	32.77 	&	42.52 	&	63.52 	&	63.57 	\\
Liver-CT	&	37.56 	&	68.32 	&	69.67 	&	68.89 	&	89.11 	&	88.28 	&	47.11 	&	70.14 	&	77.26 	&	82.13 	&	89.00 	&	88.74 	\\
Lung-CT	&	81.36 	&	86.05 	&	94.45 	&	93.64 	&	96.75 	&	93.89 	&	89.35 	&	89.59 	&	95.65 	&	95.75 	&	96.73 	&	96.40 	\\
Lung Tumor-CT	&	37.19 	&	69.33 	&	71.13 	&	75.40 	&	84.21 	&	85.02 	&	42.00 	&	64.86 	&	73.29 	&	78.44 	&	84.15 	&	84.87 	\\
Pancreas-CT	&	12.75 	&	45.88 	&	46.56 	&	51.81 	&	76.86 	&	76.38 	&	18.30 	&	39.59 	&	53.37 	&	65.77 	&	75.86 	&	75.59 	\\
Rib-CT	&	50.52 	&	74.23 	&	72.23 	&	74.97 	&	88.21 	&	86.76 	&	58.15 	&	74.07 	&	75.08 	&	79.50 	&	86.70 	&	85.80 	\\
Right Ventricle-CT	&	13.22 	&	36.66 	&	36.97 	&	47.34 	&	82.56 	&	81.67 	&	20.26 	&	38.03 	&	42.49 	&	57.26 	&	82.45 	&	81.56 	\\
Spleen-CT	&	39.11 	&	70.78 	&	72.24 	&	76.82 	&	92.54 	&	92.35 	&	50.16 	&	73.31 	&	78.36 	&	82.30 	&	92.28 	&	92.16 	\\
Stomach-CT	&	27.04 	&	52.09 	&	54.87 	&	61.87 	&	83.91 	&	84.41 	&	34.58 	&	54.46 	&	65.40 	&	74.34 	&	83.96 	&	84.89 	\\
Vertebra-CT	&	43.22 	&	61.53 	&	67.10 	&	69.64 	&	80.03 	&	76.85 	&	55.32 	&	61.40 	&	69.82 	&	73.53 	&	79.35 	&	76.63 	\\
Arota-MRI	&	75.49 	&	81.44 	&	81.09 	&	84.39 	&	90.86 	&	90.30 	&	80.86 	&	83.28 	&	84.83 	&	88.38 	&	90.55 	&	90.25 	\\
Femur-MRI	&	71.23 	&	92.60 	&	93.47 	&	92.47 	&	95.18 	&	94.94 	&	82.02 	&	92.45 	&	93.64 	&	93.71 	&	94.95 	&	94.86 	\\
Gall Bladder-MRI	&	39.84 	&	66.98 	&	65.51 	&	72.68 	&	87.97 	&	87.23 	&	52.64 	&	68.64 	&	73.24 	&	80.12 	&	87.49 	&	87.05 	\\
Kidney-MRI	&	82.30 	&	88.83 	&	88.72 	&	89.72 	&	93.65 	&	93.28 	&	87.19 	&	88.79 	&	90.05 	&	91.30 	&	93.26 	&	93.06 	\\
Liver-MRI	&	50.32 	&	82.33 	&	82.48 	&	84.07 	&	90.72 	&	90.42 	&	71.72 	&	87.24 	&	89.16 	&	89.91 	&	91.40 	&	91.31 	\\
Pancreas-MRI	&	15.02 	&	49.31 	&	48.84 	&	58.85 	&	79.73 	&	79.63 	&	27.49 	&	50.79 	&	61.97 	&	71.71 	&	79.23 	&	78.94 	\\
Prostate-MRI	&	19.09 	&	74.69 	&	76.90 	&	81.32 	&	92.85 	&	92.27 	&	40.60 	&	74.66 	&	77.93 	&	83.66 	&	92.11 	&	91.18 	\\
Spine-MRI	&	35.73 	&	64.82 	&	70.37 	&	75.04 	&	80.32 	&	80.61 	&	39.68 	&	63.79 	&	74.58 	&	77.41 	&	81.04 	&	81.96 	\\
Spleen-MRI	&	55.43 	&	81.00 	&	81.56 	&	86.62 	&	93.74 	&	93.24 	&	67.80 	&	82.66 	&	85.60 	&	87.94 	&	93.04 	&	92.68 	\\
Stomach-MRI	&	22.63 	&	54.05 	&	56.00 	&	63.72 	&	82.62 	&	82.71 	&	29.71 	&	52.76 	&	64.35 	&	73.24 	&	81.86 	&	82.22 	\\
Tibia-MRI	&	80.66 	&	93.95 	&	94.32 	&	93.75 	&	96.11 	&	95.98 	&	88.56 	&	94.01 	&	94.69 	&	94.76 	&	96.25 	&	96.13 	\\
Brain-T1W MRI	&	96.44 	&	96.62 	&	99.35 	&	98.62 	&	99.65 	&	99.49 	&	98.77 	&	98.48 	&	99.36 	&	99.39 	&	99.57 	&	99.54 	\\
Brain Tumor-T1W MRI	&	22.22 	&	38.45 	&	39.09 	&	48.68 	&	72.29 	&	71.96 	&	25.53 	&	40.24 	&	47.29 	&	55.49 	&	71.50 	&	71.92 	\\
Brain Tumor-T2W MRI	&	24.16 	&	47.59 	&	48.94 	&	57.78 	&	75.67 	&	75.33 	&	30.71 	&	49.38 	&	57.92 	&	64.09 	&	75.46 	&	75.59 	\\
Prostate-ADC MRI	&	19.47 	&	49.26 	&	50.44 	&	57.03 	&	77.03 	&	76.53 	&	34.45 	&	46.62 	&	51.21 	&	59.03 	&	75.00 	&	74.80 	\\
Left Ventricle-Cine-MRI	&	53.19 	&	82.25 	&	92.76 	&	88.62 	&	92.45 	&	93.07 	&	63.99 	&	80.66 	&	92.87 	&	91.32 	&	92.22 	&	92.36 	\\
Right Ventricle-CMR	&	36.34 	&	76.42 	&	75.68 	&	77.03 	&	89.72 	&	89.41 	&	51.33 	&	73.18 	&	79.74 	&	83.35 	&	89.06 	&	88.66 	\\
Brain-DW MRI	&	40.41 	&	84.32 	&	88.90 	&	86.74 	&	91.62 	&	90.97 	&	77.68 	&	86.07 	&	88.29 	&	89.13 	&	91.34 	&	91.42 	\\
Brain Tumor-T1-GD MRI	&	26.36 	&	41.18 	&	43.96 	&	52.02 	&	71.76 	&	71.16 	&	32.73 	&	42.94 	&	50.68 	&	57.11 	&	70.80 	&	70.83 	\\
Brain Tumor-T2-FLAIR MRI	&	25.00 	&	51.81 	&	52.42 	&	61.59 	&	77.74 	&	77.42 	&	35.61 	&	56.44 	&	64.73 	&	69.65 	&	77.99 	&	78.08 	\\
Thyroid Nodules-US	&	31.52 	&	66.57 	&	76.80 	&	78.54 	&	90.12 	&	90.30 	&	48.56 	&	71.52 	&	80.71 	&	83.95 	&	89.49 	&	89.84 	\\
Lung-X-ray	&	9.56 	&	93.25 	&	94.03 	&	93.23 	&	95.32 	&	95.11 	&	64.42 	&	91.96 	&	94.11 	&	94.20 	&	95.62 	&	95.65 	\\
Eye-Fundus	&	65.14 	&	99.30 	&	99.19 	&	99.08 	&	99.15 	&	99.22 	&	99.22 	&	99.24 	&	99.26 	&	99.23 	&	99.31 	&	99.28 	\\
Tool-Colonoscopy	&	45.59 	&	80.93 	&	87.27 	&	86.02 	&	90.89 	&	90.89 	&	75.55 	&	83.49 	&	92.31 	&	91.58 	&	91.47 	&	91.62 	\\
Polyp-Colonoscopy	&	49.49 	&	81.63 	&	85.63 	&	85.28 	&	89.94 	&	90.68 	&	71.29 	&	85.97 	&	90.28 	&	91.34 	&	90.97 	&	91.87 	\\
Adenocarcinoma-Histopathology	&	41.41 	&	74.31 	&	85.03 	&	84.96 	&	91.40 	&	90.73 	&	75.26 	&	86.15 	&	90.65 	&	90.89 	&	93.29 	&	93.17 	\\
Melanoma-Dermoscopy	&	47.43 	&	76.06 	&	81.52 	&	81.20 	&	87.47 	&	87.56 	&	61.26 	&	76.84 	&	81.69 	&	81.91 	&	86.67 	&	86.68 	\\
Cell-Microscopy	&	55.94 	&	84.34 	&	91.31 	&	80.64 	&	91.76 	&	91.60 	&	75.45 	&	79.66 	&	81.62 	&	82.21 	&	85.10 	&	85.03 	\\
Neural Structures-Electron Microscopy	&	54.63 	&	78.57 	&	78.86 	&	77.91 	&	86.99 	&	86.20 	&	61.54 	&	79.14 	&	80.85 	&	81.25 	&	87.70 	&	87.07 	\\
    \bottomrule
    \end{tabular}}%
  \label{tab:dice_select}%
\end{table*}%

\begin{table*}[!t]
  \centering
  \scriptsize
  \caption{Performance on selective common medical objects with different modalities in terms of HD (pixels). \textit{ViT-B} and \textit{ViT-H} represent the small and large encoders of SAM. $S_1$-$S_6$ represent different testing strategies, including everything, 1 point, 5 points, 10 points, box, and box with 1 point, respectively.}
  \resizebox{\textwidth}{!}{
    \begin{tabular}{lcccccc|cccccc}
    \hline
        \multirow{2}[0]{*}{Object-Modality} & \multicolumn{6}{c|}{ViT-B}                    & \multicolumn{6}{c}{ViT-H} \\
    \cline{2-13}          
    & $S_{1}$   & $S_{2}$   & $S_{3}$   & $S_{4}$   & $S_{5}$   & $S_{6}$   & $S_{1}$   & $S_{2}$   & $S_{3}$   & $S_{4}$   & $S_{5}$   & $S_{6}$ \\
    \cline{0-12}      
Arota-CT	&	47.70 	&	21.57 	&	18.68 	&	17.03 	&	7.50 	&	7.17 	&	39.39 	&	20.18 	&	13.62 	&	10.92 	&	7.68 	&	7.28 	\\
Brain-CT	&	37.49 	&	16.54 	&	15.96 	&	16.46 	&	8.80 	&	8.64 	&	23.34 	&	14.50 	&	12.99 	&	12.86 	&	8.47 	&	8.71 	\\
Heart-CT	&	233.23 	&	154.20 	&	135.36 	&	124.77 	&	23.49 	&	23.21 	&	227.05 	&	124.27 	&	108.98 	&	92.50 	&	23.73 	&	23.67 	\\
Humerus-CT	&	16.41 	&	4.77 	&	4.95 	&	5.42 	&	2.24 	&	2.37 	&	11.99 	&	4.86 	&	4.06 	&	3.84 	&	2.19 	&	2.21 	\\
Kidney-CT	&	40.48 	&	18.47 	&	18.43 	&	18.94 	&	9.65 	&	10.25 	&	29.20 	&	18.71 	&	15.44 	&	13.36 	&	10.37 	&	10.52 	\\
Kidney Tumor-CT	&	84.17 	&	47.15 	&	45.58 	&	39.83 	&	8.10 	&	8.47 	&	69.88 	&	36.69 	&	20.62 	&	17.90 	&	7.86 	&	7.83 	\\
Left Atrium-CT	&	81.14 	&	48.98 	&	49.74 	&	44.15 	&	4.65 	&	5.22 	&	71.31 	&	42.63 	&	40.23 	&	29.97 	&	4.66 	&	5.00 	\\
Left Ventricular Myocardium-CT	&	105.43 	&	60.31 	&	57.95 	&	42.37 	&	12.58 	&	12.65 	&	99.41 	&	56.98 	&	48.41 	&	36.98 	&	12.00 	&	12.21 	\\
Liver-CT	&	149.00 	&	85.65 	&	78.72 	&	84.59 	&	28.22 	&	29.52 	&	127.53 	&	80.90 	&	58.79 	&	50.77 	&	27.38 	&	27.92 	\\
Lung-CT	&	57.69 	&	45.90 	&	21.56 	&	42.70 	&	12.11 	&	16.76 	&	48.44 	&	39.36 	&	14.98 	&	22.63 	&	10.76 	&	10.50 	\\
Lung Tumor-CT	&	108.83 	&	34.87 	&	34.73 	&	28.32 	&	9.28 	&	9.11 	&	96.89 	&	43.34 	&	30.83 	&	22.82 	&	9.14 	&	9.07 	\\
Pancreas-CT	&	135.29 	&	52.93 	&	51.52 	&	38.48 	&	12.39 	&	12.63 	&	120.75 	&	68.38 	&	41.53 	&	23.39 	&	13.48 	&	13.50 	\\
Rib-CT	&	60.27 	&	15.81 	&	15.25 	&	10.69 	&	1.58 	&	1.67 	&	48.03 	&	16.75 	&	10.66 	&	7.36 	&	1.69 	&	1.72 	\\
Right Ventricle-CT	&	101.42 	&	61.21 	&	59.00 	&	36.99 	&	9.17 	&	9.57 	&	93.08 	&	53.03 	&	44.63 	&	31.36 	&	9.32 	&	9.80 	\\
Spleen-CT	&	125.28 	&	46.89 	&	42.79 	&	34.44 	&	6.44 	&	6.61 	&	100.26 	&	46.52 	&	33.48 	&	24.36 	&	6.37 	&	6.41 	\\
Stomach-CT	&	93.98 	&	59.79 	&	59.24 	&	41.01 	&	11.80 	&	11.90 	&	84.03 	&	57.43 	&	41.88 	&	26.79 	&	11.88 	&	11.68 	\\
Vertebra-CT	&	46.89 	&	20.71 	&	16.33 	&	15.67 	&	7.99 	&	8.68 	&	31.85 	&	21.45 	&	14.58 	&	12.37 	&	8.25 	&	8.76 	\\
Arota-MRI	&	26.19 	&	11.79 	&	11.93 	&	8.29 	&	3.42 	&	3.57 	&	17.85 	&	11.21 	&	7.56 	&	5.12 	&	3.49 	&	3.61 	\\
Femur-MRI	&	65.49 	&	20.03 	&	19.62 	&	41.44 	&	13.93 	&	14.89 	&	46.92 	&	18.33 	&	16.89 	&	20.80 	&	13.91 	&	13.31 	\\
Gall Bladder-MRI	&	109.02 	&	27.99 	&	30.96 	&	18.59 	&	5.06 	&	5.22 	&	70.11 	&	26.42 	&	17.67 	&	10.81 	&	5.15 	&	5.24 	\\
Kidney-MRI	&	28.27 	&	13.43 	&	13.36 	&	12.55 	&	7.67 	&	8.72 	&	18.35 	&	13.69 	&	11.57 	&	10.41 	&	8.38 	&	8.46 	\\
Liver-MRI	&	113.93 	&	46.71 	&	43.86 	&	51.74 	&	22.66 	&	23.54 	&	70.66 	&	37.84 	&	31.01 	&	29.66 	&	23.17 	&	23.28 	\\
Pancreas-MRI	&	123.47 	&	48.35 	&	50.86 	&	32.53 	&	11.13 	&	11.26 	&	93.84 	&	43.83 	&	25.86 	&	18.08 	&	11.62 	&	11.68 	\\
Prostate-MRI	&	195.89 	&	59.22 	&	54.08 	&	55.25 	&	17.87 	&	19.04 	&	158.55 	&	50.40 	&	43.87 	&	39.60 	&	17.10 	&	17.52 	\\
Spine-MRI	&	116.20 	&	44.59 	&	29.91 	&	21.17 	&	13.67 	&	13.32 	&	111.57 	&	46.92 	&	26.02 	&	19.66 	&	13.46 	&	12.28 	\\
Spleen-MRI	&	90.79 	&	28.53 	&	27.65 	&	17.61 	&	5.87 	&	6.25 	&	61.74 	&	23.78 	&	17.26 	&	14.83 	&	6.06 	&	6.32 	\\
Stomach-MRI	&	126.55 	&	52.28 	&	56.81 	&	38.46 	&	12.72 	&	12.89 	&	107.99 	&	53.84 	&	37.97 	&	25.21 	&	13.35 	&	13.41 	\\
Tibia-MRI	&	49.40 	&	14.14 	&	19.30 	&	48.64 	&	7.64 	&	8.41 	&	27.62 	&	11.32 	&	10.65 	&	14.27 	&	6.76 	&	6.88 	\\
Brain-T1W MRI	&	7.16 	&	7.26 	&	3.51 	&	8.70 	&	2.80 	&	3.13 	&	4.90 	&	5.96 	&	3.66 	&	3.30 	&	2.62 	&	2.70 	\\
Brain Tumor-T1W MRI	&	78.08 	&	60.05 	&	62.43 	&	46.91 	&	15.35 	&	16.13 	&	74.02 	&	55.87 	&	47.83 	&	35.79 	&	15.16 	&	14.83 	\\
Brain Tumor-T2W MRI	&	75.45 	&	49.85 	&	50.90 	&	35.79 	&	13.61 	&	14.07 	&	67.80 	&	46.41 	&	37.09 	&	27.97 	&	13.85 	&	13.83 	\\
Prostate-ADC MRI	&	89.48 	&	43.56 	&	43.26 	&	37.06 	&	12.85 	&	13.26 	&	76.12 	&	45.17 	&	39.73 	&	31.00 	&	13.83 	&	13.93 	\\
Left Ventricle-Cine-MRI	&	33.68 	&	6.68 	&	3.39 	&	7.37 	&	3.23 	&	3.14 	&	24.80 	&	6.61 	&	3.18 	&	3.97 	&	3.21 	&	3.27 	\\
Right Ventricle-CMR	&	78.26 	&	18.12 	&	19.76 	&	21.19 	&	4.85 	&	5.08 	&	63.36 	&	20.81 	&	12.98 	&	11.05 	&	4.96 	&	5.10 	\\
Brain-DW MRI	&	49.27 	&	26.48 	&	18.97 	&	24.72 	&	13.88 	&	15.81 	&	28.20 	&	23.32 	&	19.66 	&	17.83 	&	14.42 	&	14.48 	\\
Brain Tumor-T1-GD MRI	&	69.26 	&	51.83 	&	54.06 	&	41.86 	&	15.07 	&	15.77 	&	60.97 	&	48.60 	&	41.69 	&	31.85 	&	15.05 	&	15.10 	\\
Brain Tumor-T2-FLAIR MRI	&	75.20 	&	46.69 	&	48.16 	&	33.41 	&	13.02 	&	13.45 	&	63.10 	&	39.36 	&	30.26 	&	23.24 	&	13.06 	&	13.08 	\\
Thyroid Nodules-US	&	163.70 	&	77.62 	&	63.03 	&	62.41 	&	20.86 	&	21.61 	&	127.29 	&	63.00 	&	42.05 	&	34.78 	&	20.71 	&	20.53 	\\
Lung-X-ray	&	1977.68 	&	257.39 	&	236.67 	&	448.30 	&	159.31 	&	175.42 	&	829.27 	&	272.97 	&	210.19 	&	251.03 	&	129.96 	&	126.08 	\\
Eye-Fundus	&	147.05 	&	84.68 	&	7.75 	&	94.83 	&	4.56 	&	37.85 	&	4.28 	&	19.03 	&	5.47 	&	4.26 	&	3.86 	&	3.97 	\\
Tool-Colonoscopy	&	326.06 	&	150.49 	&	102.04 	&	169.69 	&	49.88 	&	52.50 	&	189.25 	&	128.34 	&	43.01 	&	63.74 	&	43.45 	&	43.77 	\\
Polyp-Colonoscopy	&	214.44 	&	111.48 	&	101.82 	&	110.13 	&	59.50 	&	60.51 	&	145.66 	&	96.13 	&	78.36 	&	78.91 	&	50.53 	&	50.15 	\\
Adenocarcinoma-Histopathology	&	299.05 	&	89.15 	&	63.26 	&	93.00 	&	30.33 	&	32.58 	&	126.99 	&	55.09 	&	41.77 	&	46.82 	&	23.41 	&	24.51 	\\
Melanoma-Dermoscopy	&	947.16 	&	444.84 	&	417.83 	&	538.32 	&	274.31 	&	283.00 	&	697.27 	&	397.32 	&	359.48 	&	352.98 	&	259.81 	&	262.04 	\\
Cell-Microscopy	&	82.76 	&	17.00 	&	13.30 	&	28.95 	&	6.93 	&	7.67 	&	33.55 	&	24.61 	&	22.84 	&	23.29 	&	16.29 	&	16.37 	\\
Neural Structures-Electron Microscopy	&	126.14 	&	20.55 	&	18.80 	&	30.54 	&	7.52 	&	8.40 	&	95.13 	&	19.65 	&	14.19 	&	19.14 	&	6.97 	&	7.22 	\\
    \bottomrule
    \end{tabular}}%
  \label{tab:hd_select}%
\end{table*}%

\subsection{Segmentation Performance under Different Testing Modes}

In this section, we tend to compare the segmentation performance among different strategies using different models (\textit{ViT-B} and \textit{ViT-H}).
As shown in Fig.~\ref{fig:strategy_dice_hd}, we show the average DICE and HD performance of \textit{ViT-B} and \textit{ViT-H} under different strategies.
For both \textit{ViT-B} and \textit{ViT-H}, the performance trends of different strategies are basically consistent.
\textit{Everything} ($S_1$) achieve the worst performance.
For the point prompts ($S_2$-$S_4$), adding more points will bring stable performance improvement (ViT-B: DICE from 56.02\% to 63.81\%, ViT-H: DICE from 56.78\% to 71.61\%).
SAM with a box prompt yields the best performance, while adding one point to the box will not bring obvious changes (ViT-B: DICE 0.49\%$\downarrow$, ViT-H: DICE 0.06\%$\downarrow$).
Based on the experiments, we conclude that box prompts include more vital information compared to point prompts.
Since the box actually tells the exact location of the target and also its potential intensity features given the limited region.
However, points only represent the part features of the target, which can lead to confusion.
Fig.~\ref{fig:choose_b}, Fig.~\ref{fig:choose_h}, Table~\ref{tab:dice_select} and Table~\ref{tab:hd_select} present specific segmentation accuracy results for partial targets under six testing strategies.
See Fig.~\ref{fig:good_data} for prediction visualization of the 5 manual prompting modes ($S_2$-$S_6$). For additional results, please refer to the supplementary materials.

\begin{table}[!t]
  \centering
  \caption{Test time analysis of SAM (seconds).}
   \resizebox{0.48\textwidth}{!}{
    \begin{tabular}{cccccccc}
    \toprule
         \multirow{2}[4]{*}{Model}  & \multirow{2}[4]{*}{Embedding} & \multicolumn{6}{c}{Prompt Encoding+Mask Decoding} \\
\cmidrule{3-8}          &       & $S_{1}$ & $S_{2}$ & $S_{3}$ & $S_{4}$ & $S_{5}$ & $S_{6}$ \\
    \midrule
     ViT-B & 0.1276  & 1.9692  & 0.0085  & 0.0088  & 0.0090  & 0.0080  & 0.0088  \\
    ViT-H & 0.4718  & 3.0324  & 0.0086  & 0.0088  & 0.0091  & 0.0080  & 0.0090  \\
    \bottomrule
    \end{tabular}%
  \label{tab:sam_time}}%
\end{table}%

\begin{figure*}[!ht]
	\centering
	\includegraphics[width=1.0\linewidth]{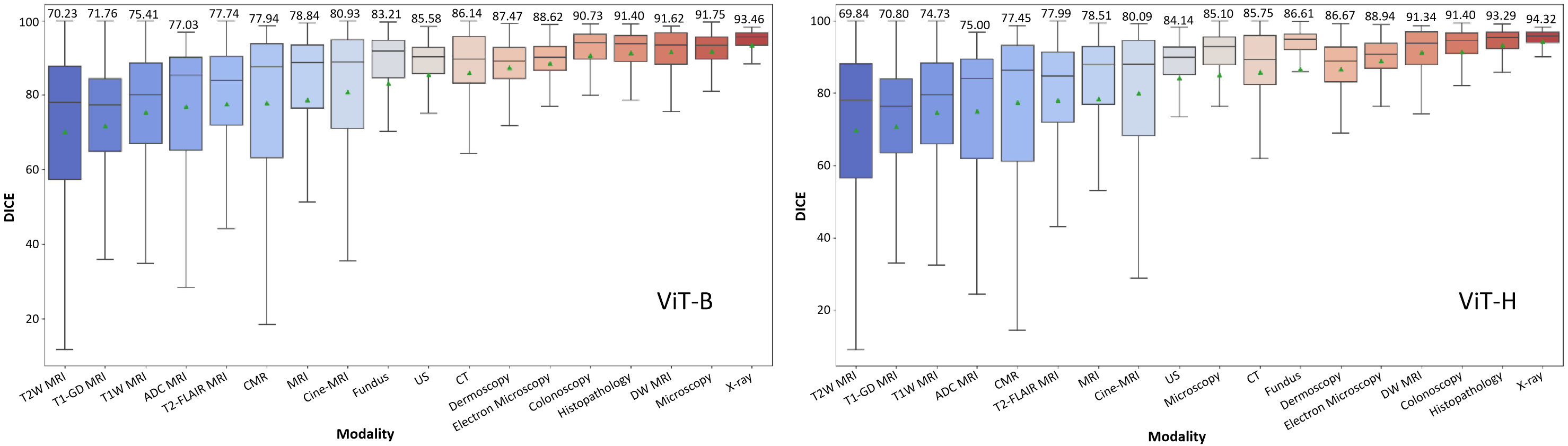}
	\caption{DICE performance of 18 different modalities. The green triangle and the values above each box in the box plot represent the average value.}
	\label{fig:single_modality}
\end{figure*}

\subsection{Performance of Medical Images on Different Modalities}

In Fig.~\ref{fig:single_modality}, we summarized SAM's performance under box prompts for different modalities. 
SAM shows the highest average DICE performance on the X-ray modality for both models (\textit{ViT-B} and \textit{ViT-H}).
In modalities including histopathology, colonoscopy, and DW MRI, SAM also achieved satisfactory DICE performance ($>$90\%).
Additionally, their standard deviations are relatively low.
This proves that SAM performs well and stably in the objects of these modalities.
There are 6 (\textit{ViT-B}) and 7 (\textit{ViT-H}) modalities with DICE performance between 80 and 90, but most of them have large standard deviations.
SAM performs worse in the remaining modalities, and the results are also more unstable.
Note that the same objects with different modalities may achieve slightly varying segmentation performance.
See \textit{BrainTumor} in Table~\ref{tab:dice_select} and~\ref{tab:hd_select}, its performance ranges from 71.76\% to 77.74\% in DICE and 13.02 pixels to 15.35 pixels in HD.
Although segmentation performance can be affected by various factors, we hope that Fig.~\ref{fig:single_modality} could provide basic guidance for researchers on how to use SAM appropriately for different modalities.

\subsection{Inference Time Analysis of SAM}

Inference time is an important factor in evaluating a model.
In Table~\ref{tab:sam_time}, we reported the average inference time in terms of embedding generation, prompt encoding, and mask decoding.
All the testing is performed on one NVIDIA GTX 3090 GPU with 24G memory.
Testing time may be influenced by lots of factors, including the image size (tiny difference in pre-processing time, i.e., upsampling the images with different sizes to 1024$\times$1024), and the number of targets (processing the prompts for each target in serial).
Thus, we conducted a fair comparison by limiting the image size to 256$\times$256 and setting the number of targets to 1.
It can be observed that \textit{ViT-H} takes nearly four times the embedding time of \textit{ViT-B}.
Prompt encoding and mask decoding for \textit{Everything} ($S_{1}$) is extremely time-consuming, as it requires processing hundreds of points sampled from the whole image, including heavy post-processing using NMS, etc.
While for the manual prompt encoding and mask decoding ($S_{2}$-$S_{6}$), the reference times of different models and strategies are similar and less than 0.01s.
We believe that the evaluation time of SAM in manual mode can meet the needs of real-time use.

\subsection{Analysis on the Number of Points in \textit{Everything} Mode}
\begin{figure}[!ht]
	\centering
	\includegraphics[width=1 \linewidth]{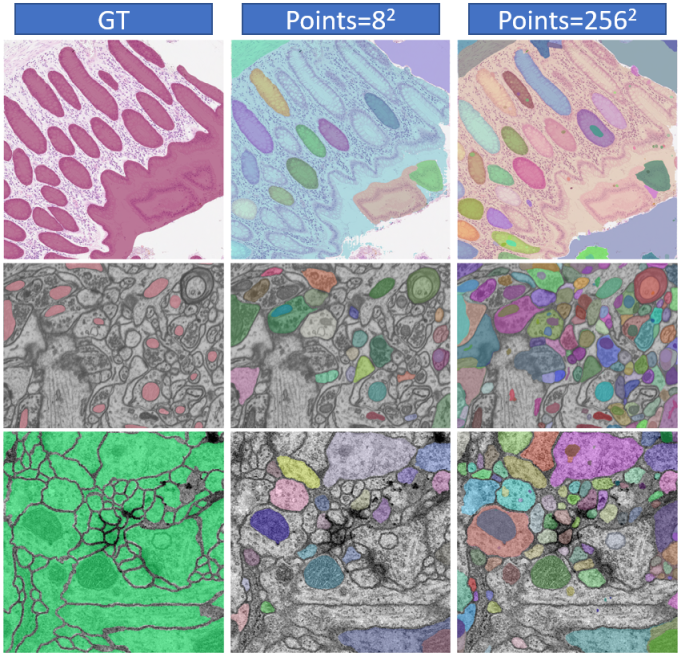}
	\caption{Different cases of Adenocarcinoma~\citep{sirinukunwattana2017gland}, Mitochondria~\citep{lucchi2013learning} and Neural Structures~\citep{cardona2010integrated} with different numbers of points in $S_{1H}$.}
	\label{fig:everything}
\end{figure}

As described above, in the \textit{Everything} mode, a grid of point prompts (m$\times$m) will be generated.
In default, m is set to 32.
The number of points will have an impact on the final segmentation performance.
Especially for images that have multiple targets with different sizes, improper parameter designs will lead to imperfect segmentation with some objects being unprompted.
As shown in Table~\ref{tab:points_everything}, we tested four datasets with multiple objects on one image.
The results show that in these four datasets, as the number of points increases from $8^2$ to $256^2$, the DICE also increases gradually.
Fig.~\ref{fig:everything} also shows that more points will bring more potential objects (shown in different colors).
Additionally, too many points make SAM split an object into several pieces, destroying the integrity of the object.
Increasing the number of points can also lead to a significant increase in test time.
Hence, it is a trade-off between segmentation performance and test efficiency.

\begin{table}[htbp]
	\centering
	\caption{Ablation study on the number of points in \textit{Everything} mode.}
	\resizebox{0.48\textwidth}{!}{
		\begin{tabular}{ccccccc}
			\toprule
		 {Objects}	& $8^2$     & $16^2$    & $32^2$    & $64^2$    & $128^2$   & $256^2$ \\
			\midrule
			Adenocarcinoma-Histopathology~\citep{sirinukunwattana2017gland} &  $42.3_{44.1}$  &  $70.4_{37.4}$   &  $76.3_{32.5}$   &  $77.8_{31.1}$     &    $78.7_{30.4}$   &  $79.1_{30.2}$ \\
			Adenocarcinoma-Histopathology~\citep{graham2019mild} &  $55.6_{42.4}$   &  $70.9_{34.7}$  &  $74.2_{31.5}$   &  $76.3_{30.0}$  &  $77.1_{29.1}$  &  $77.6_{28.6}$\\
			Mitochondria-Electron Microscopy~\citep{lucchi2013learning} &   $31.8_{39.3}$    & $71.1_{31.4}$      &  $81.1_{20.1}$     &  $81.5_{19.1}$     &   $81.7_{18.6}$    &  $81.7_{18.3}$\\
			Neural Structures-Microscopy~\citep{cardona2010integrated} &  $21.3_{38.0}$    &  $45.7_{44.1}$     &  $61.5_{40.1}$     &  $64.4_{38.4}$     &   $65.4_{37.8}$    & $65.8_{37.5}$  \\
			\bottomrule
	\end{tabular}}%
	\label{tab:points_everything}%
\end{table}%

\begin{figure}[!ht]
	\centering
	\includegraphics[width=1\linewidth]{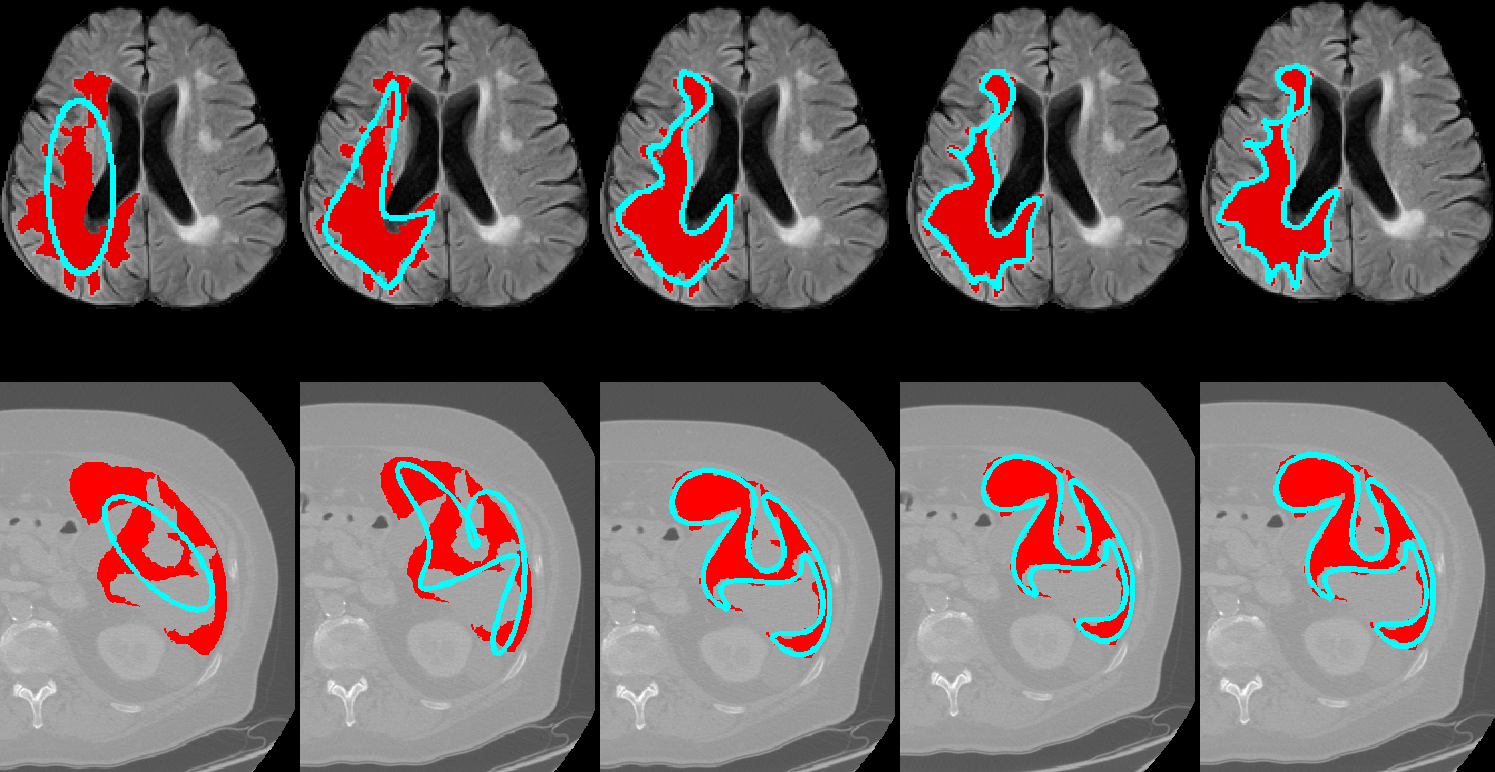}
	\caption{Contour decoded from the Fourier series. From left to right, the decoded contour (blue) gets closer to the original contours (red) as the FO increases.}
	\label{fig:fourier_grow}
\end{figure}

\subsection{Analysis of Factors Correlating to Segmentation Results}
To verify the factors that affect the segmentation performance of SAM, we recorded the size, aspect ratio, intensity difference between foreground and background, modality, and boundary complexity of 191,779 anatomical structures. 
By analyzing these factors, we aim to better understand the correlation between anatomical structure characteristics and SAM's segmentation performance, and further provide some useful insights into the development of medical SAM.\par

The size of the anatomical structure was computed as the pixel-level area of the corresponding mask. To determine the aspect ratio of a mask, we need to calculate the ratio (ranging from 0 to 1) between the short and long sides of its bounding box.
The intensity difference was defined as the variation in mean intensity values between the structure and the surrounding area within an enlarged bounding box, excluding the structure itself. Specifically, to accommodate the varied dimensions of targets, we dynamically expanded the box outward by a preset ratio of $0.1$, instead of using fixed pixel values (e.g., extending by 10 pixels).
Additionally, the modality of each anatomical structure was mapped to numerical values.
Furthermore, we introduced Elliptical Fourier Descriptors (EFD)~\citep{EFD} to describe boundary complexity.


\begin{table}[!t]
	\centering
         \scriptsize
	\caption{Spearman partial correlation analysis (values with $p<0.001$ are shown in bold).}
	\resizebox{.5\textwidth}{!}{
		\begin{tabular}{c|cccccccccc}
			\toprule
			\multirow{2}[0]{*}{Strategy} & \multicolumn{2}{c}{Size } & \multicolumn{2}{c}{Intensity Difference} &  \multicolumn{2}{c}{Fourier Order} & \multicolumn{2}{c}{Modality}& \multicolumn{2}{c}{Aspect Ratio}\\
            \cline{2-11}
           & ViT-B  & ViT-H  & ViT-B  & ViT-H & ViT-B  & ViT-H & ViT-B  & ViT-H & ViT-B  & ViT-H\\
			\cline{0-10}
                $S_{1}$&0.218&0.271&\textbf{0.503}&\textbf{0.614}&\textbf{-0.419}&\textbf{-0.453}&-0.006&0.048&0.056&0.035\\
                $S_{2}$& 0.236&0.293& \textbf{0.572}&\textbf{0.535}& \textbf{-0.537}& \textbf{-0.519}& 0.051& 0.049&0.079&0.072\\
                $S_{3}$&0.289&0.330&\textbf{0.638}&\textbf{0.591}&\textbf{-0.520}&\textbf{-0.537}&0.094&0.065&0.046&0.053\\
                $S_{4}$&0.275&0.339&\textbf{0.628}&\textbf{0.533}&\textbf{-0.524}&\textbf{-0.533}&0.062&0.040&0.048&0.060\\
			$S_{5}$&  \textbf{0.410} & \textbf{0.428}&\textbf{0.445}&\textbf{0.407}&\textbf{-0.479}	&\textbf{-0.463}&-0.023&-0.034&0.065&0.076 \\
            $S_{6}$&0.370&0.392&\textbf{0.467}&\textbf{0.412}&\textbf{-0.621}&\textbf{-0.576}&0.014&-0.011&0.068&0.071\\
			\bottomrule
	\end{tabular} }%
	\label{tab:partical}%
\end{table}%

\begin{figure*}[!t]
	\centering
	\includegraphics[width=1.0\linewidth]{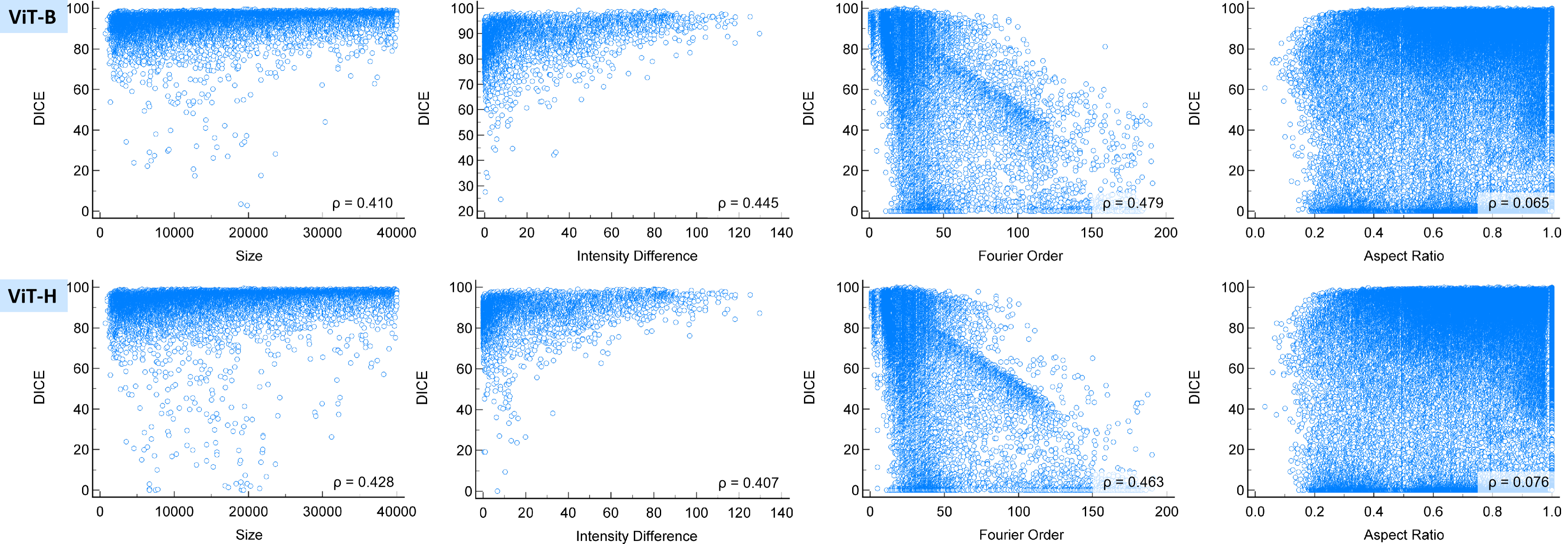}
	\caption{Scatterplot of different object attributes with DICE under $S_{5}$ strategies.}
	\label{fig:icc_Scatterplot}
\end{figure*}

\begin{figure}[!t]
	\centering
	\includegraphics[width=1.0\linewidth]{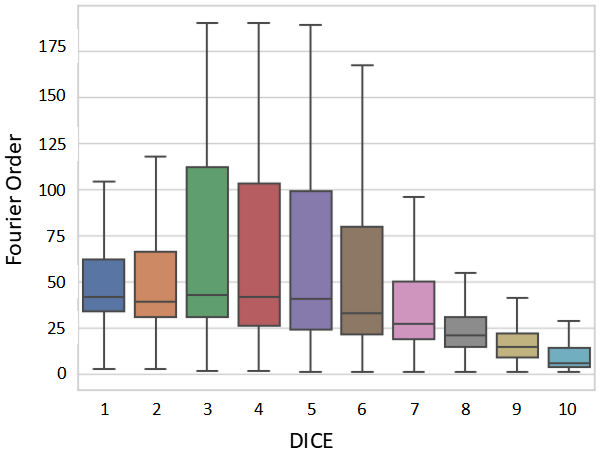}
	\caption{FO boxplot of different DICE ranges. }
	\label{fig:icc_box_plot}
\end{figure}

EFD encodes the contour of a mask into a Fourier series that represents different frequency components. As the Fourier order (FO) increases, the contour decoded from the Fourier series gets closer to the original contour (see Fig.~\ref{fig:fourier_grow}), and the decoding process can be described as Eq.~\ref{qua:fourier}. 
\begin{equation}
	\label{qua:fourier}
	\begin{matrix}
		x_{N}(t) = L_x + \sum_{n=1}^N (a_n\sin{(\frac{T}{2n\pi t}} + b_n\cos{(\frac{T}{2n\pi t})})\\
		\vspace{0.05em}\\
		y_{N}(t) = L_y + \sum_{n=1}^N (c_n\sin{(\frac{T}{2n\pi t}} + d_n\cos{(\frac{T}{2n\pi t})})
	\end{matrix},
\end{equation}
\vspace{0.05em}\\
where $(x_{N}(t), y_{N}(t))$ is the coordinates of any point on the contour, $N$ is the number of Fourier series expansions, and $t\in [0, T]$ denotes the different sampling locations. $(L_x, L_y)$ indicates the coordinates of the contour’s center point, $(a_n, b_n)$ denotes the parameter obtained by Fourier coding of the x-coordinates, and $(c_n, d_n)$ denotes the encoding result in the y-direction.
We can roughly estimate the complexity of the object boundary in terms of the FO.
Specifically, the order is defined as the required number of accumulations when the contour from the decoded Fourier series reaches a certain degree of overlap (we use DICE to represent overlap) with the original contour.
However, when using this approach as a quantitative measure, it is especially important to set an appropriate DICE threshold. 

A low threshold cannot accurately distinguish the difference in complexity among various object boundaries.
If the threshold is too high, the EFD may fail to fit the complex contours as required and get into an infinite calculation.
Thus, we optimized the representation of FO to avoid the EFD program from getting stuck in endless accumulation (see cumulative terms in Eq.~\ref{qua:fourier}).

For different structures, we increased the FO from 1 and calculated the DICE between the decoded contour and the original contour at each step.
Then we set two ways to end the process: 1) $DICE >97.0 \%$; 2) the difference in the DICE between order $F_{(a-1)}$ and order $F_{(a)}$ is less than $0.1 \%$. 
Consequently, we record the FO ($F_{(a)}$) and DICE after termination. 
Finally, we take $F_{final} = F_{a} + n\times100\times(1-DICE), n=2$ as the final optimized FO.\par

\begin{figure}[!t]
	\centering
	\includegraphics[width=1.0\linewidth]{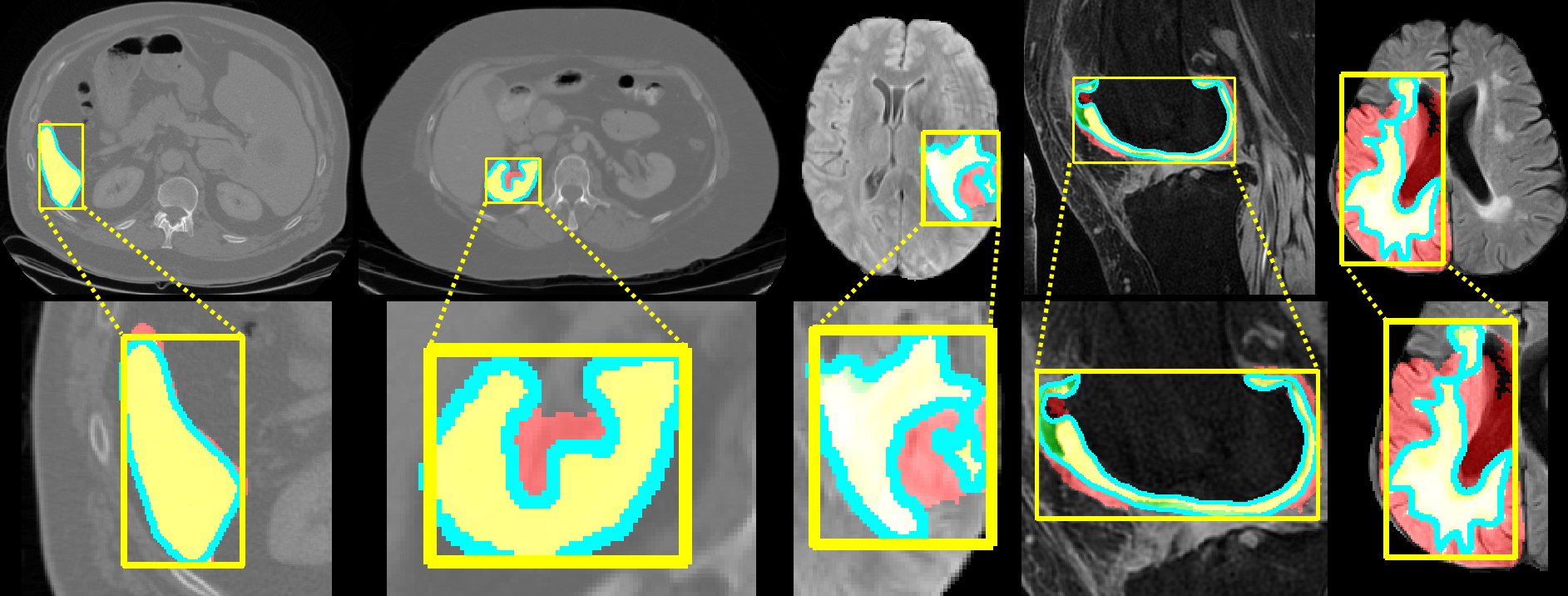}
	\caption{Relationship between DICE and FO. From left to right, FO gradually increases. The yellow box represents the box prompt, the red mask is the prediction, the green mask is the GT, the yellow mask is the overlap of prediction and GT, and the blue contour is decoded from the Fourier series.}
	\label{fig:icc_visual}
\end{figure}
We conducted an analysis of the partial correlation between the five attributes of the target objects mentioned above and the DICE score using Spearman's rank partial correlation coefficient under different testing strategies.
The statistical results are displayed in Table~\ref{tab:partical}, while Fig.~\ref{fig:icc_Scatterplot} illustrates the scatterplot for the $S_{5}$ strategies. Across most testing strategies, we observed that the DICE score exhibited a moderate correlation (\textit{$0.4 \leq \rho < 0.7$}) with the FO and intensity difference, a weak correlation (\textit{$0.2 \leq \rho < 0.4$}) with the size, and no correlation with the modality and aspect ratio.
Thus, SAM can consistently segment medical targets with different modality and aspect ratios.
The performance of SAM under box prompt might be affected by the size of anatomical structures.
Furthermore, the performance of SAM tends to be poor under all testing strategies when dealing with objects characterized by complex boundaries or low contrast.
To confirm these findings, we divided DICE calculated under $S_{5B}$ into ten levels on average (\textit{e.g.,} level-1 means DICE (\%) belongs to (0,10]) and visualized the FO boxplot for different DICE levels in Fig.~\ref{fig:icc_box_plot}. 
The figure illustrates that as the DICE level increases, the FO distribution of the structures gradually shifts to a range with smaller values.
Additionally, in Fig.~\ref{fig:icc_visual}, we presented visualizations of anatomical structures with various FO ranges. These visualizations reveal that the DICE score of anatomical structures tends to decrease as the FO increases. It further implies that shape and boundary complexity may have an impact on SAM's segmentation performance.

\subsection{Annotation Time and Quality Analysis}

In this section, we discuss whether the SAM can help doctors improve annotation time and quality.
We randomly sampled 100 images with average DICE performance from \textit{\ourdata}, to construct an evaluation subset comprising 55 objects and 620 masks across 9 modalities, including instances of the same object in different modalities. 
We then invited three doctors with 10 years of experience to evaluate whether SAM's prediction under box prompts could improve annotation speed and quality. 
They were given tasks including 1) annotating all objects in the evaluation subset from scratch, 2) adjusting object labels based on SAM's predictions, and 3) recording the time for both tasks. 
To evaluate annotation quality, we utilize the Human Correction Efforts (HCE) index~\citep{qin2022highly}, which estimates the human effort required to correct inaccurate predictions to meet specific accuracy (i.e., GT masks) requirements in real-world applications.
The lower HCE index represents that the mask (annotation of human with/without SAM) is closer to GT, i.e., annotation is of higher quality.
As shown in Table~\ref{tab:user_val}, with SAM's help, it can attain higher annotation quality (HCE: 0.27$\downarrow$) and boost annotation speed by approximately 25\%.
Specifically, $\sim$1.31 minutes can be saved for annotating one image, and $\sim$0.2 minutes for one object (since one image contains $\sim$6.2 objects in the above task).
The greater the number of anatomical structures that need to be labeled, the more obvious the advantage of SAM's efficiency will be.

\begin{table}[!t]
  \centering
  \caption{Annotation speed and quality of a human with or without SAM's help. s: seconds, m: minutes.}
  \resizebox{0.48\textwidth}{!}{
    \begin{tabular}{cc|c|cccc}
    \toprule
    \multicolumn{2}{c|}{SAM} & \multirow{2}[1]{*}{Doctor} & \multicolumn{2}{c}{Human } & \multicolumn{2}{c}{Human with SAM} \\
    \cline{0-1}\cline{4-5}\cline{6-7}
    HCE$\downarrow$   & Time (s) &       & HCE$\downarrow$   & Time (m) & HCE$\downarrow$   & Time (m) \\
    \cline{0-6}
    \multirow{4}[1]{*}{5.66} & \multirow{4}[1]{*}{0.47} & Doctor1 & 4.74  & 4.41  & \textbf{4.57 } & \textbf{3.03 } \\
          &      & Doctor2 & 5.82  & 4.21  & \textbf{5.23 } & \textbf{2.95 } \\
          &       & Doctor3 & 4.65  & 4.19  & \textbf{4.59 } & \textbf{2.91 } \\
          &       & Mean  & 5.07  & 4.27  & \textbf{4.80 } & \textbf{2.96 } \\
    \bottomrule
    \end{tabular}}%
  \label{tab:user_val}%
\end{table}%

\subsection{Impact of Different Prompt Randomness on Performance}
In the previous experiments, we fixed the box and point selection strategy for the repeatability of the experiments.
We tested the theoretical optimum performance of SAM via the selection of the center of mass and tight box because they may include the most representative features of the target.
However, it is not practical to click the exact center or draw the exact box of each object to evaluate SAM.
Thus, we added different levels of randomness to the centers and boxes to simulate real-life human operation~\citep{huang2021flip}.
Besides, we believe that this can help us better discuss the robustness of SAM.

Specifically, we enlarge/move the boxes/points randomly in 0-10, 10-20, and 20-30 pixels.
In Table~\ref{tab:randomness}, the random experiments (Random 1-3) were conducted three times, and the average results were calculated (Mean).
\textit{DICE drop} represents the average declining DICE value compared to the original results without shifting.
For $S_2$ (single point), the DICE performance dropped by 2.67\%, 7.38\%, and 14.62\% as the shifting level increased. 
With the increase in the number of point prompts ($S_3$ and $S_4$), the decline of DICE could be alleviated, and the stability of the model could be improved.
SAM was severely affected by box offsets ($S_5$, 24.11\% decrease in performance for shifts of 20-30 pixels), while this impact was even more pronounced when adding one point to the box ($S_6$, with a decrease of 29.93\%).

\begin{table}[!h]
  \centering
  \scriptsize
  \caption{Comparison of DICE decrease under different shifting levels and testing strategies.}
  \resizebox{0.48\textwidth}{!}{
    \begin{tabular}{c|c|ccccc}
    \toprule
          & \multirow{2}[0]{*}{Shift} & \multicolumn{5}{c}{DICE drop} \\
\cline{3-7}          &       & $S_2$    &  $S_3$    &  $S_4$    &  $S_5$    &  $S_6$ \\
    \cline{0-6}
    \multirow{3}[2]{*}{Random 1} & 0-10  & 2.74  & 0.87  & 0.79  & 3.08  & 4.57  \\
          & 10-20 & 7.55  & 1.42  & 1.37  & 10.24  & 13.98  \\
          & 20-30 & 14.36  & 4.67  & 3.29  & 23.88  & 29.72  \\
    \cline{0-6}
    \multirow{3}[2]{*}{Random 2} & 0-10  & 2.68  & 0.90  & 0.84  & 3.24  & 4.49  \\
          & 10-20 & 7.42  & 1.38  & 1.29  & 10.39  & 13.83  \\
          & 20-30 & 14.51  & 4.49  & 3.17  & 23.71  & 29.50  \\
    \cline{0-6}
    \multirow{3}[2]{*}{Random 3} & 0-10  & 2.60  & 0.81  & 0.73  & 3.43  & 5.02  \\
          & 10-20 & 7.17  & 1.19  & 1.22  & 10.87  & 14.28  \\
          & 20-30 & 14.98  & 4.15  & 3.05  & 24.73  & 30.58  \\
    \cline{0-6}
    \multirow{3}[2]{*}{Mean} & 0-10  & 2.67  & 0.86  & 0.79  & 3.25  & 4.69  \\
          & 10-20 & 7.38  & 1.33  & 1.29  & 10.50  & 14.03  \\
          & 20-30 & 14.62  & 4.44  & 3.17  & 24.11  & 29.93  \\
    \bottomrule
    \end{tabular}}%
  \label{tab:randomness}%
\end{table}%

\begin{figure}[!ht]
	\centering
	\includegraphics[width=1\linewidth]{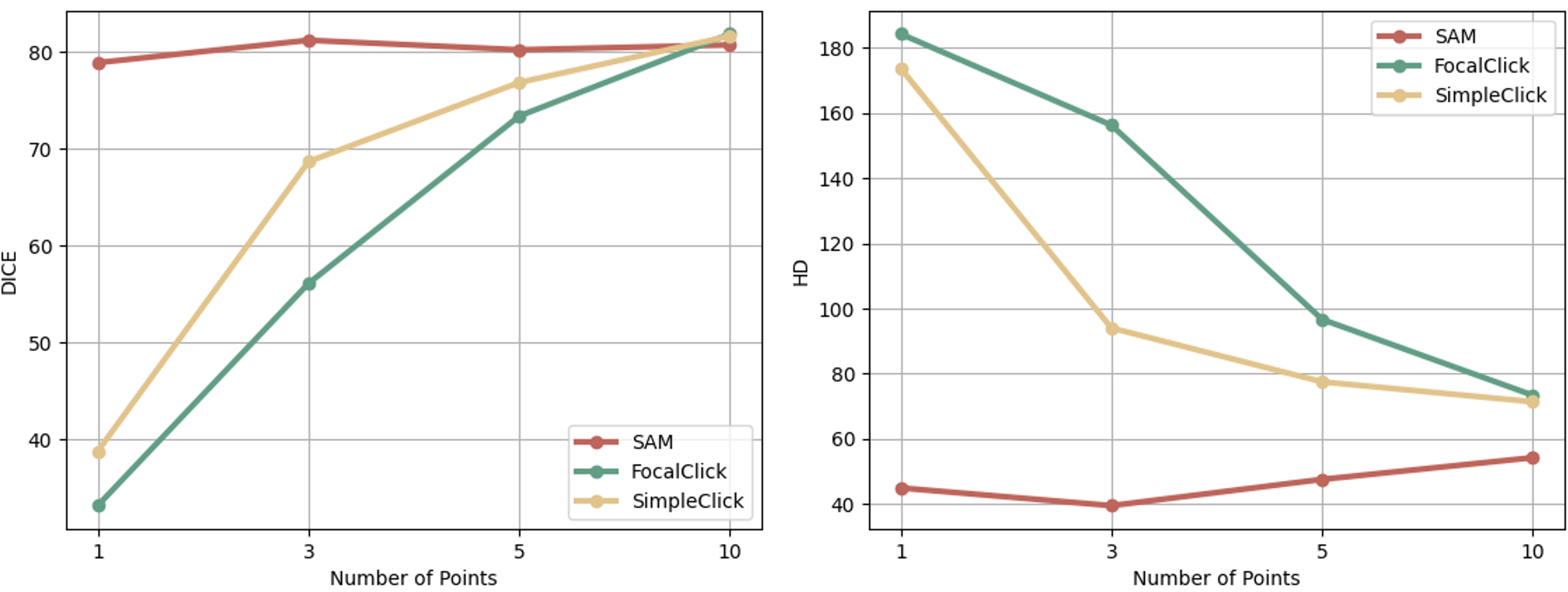}
	\caption{Average performance of three different methods varies with the number of point prompts.}
	\label{fig:interaction}
\end{figure}

\begin{figure*}[!ht]
	\centering
	\includegraphics[width=1.0\linewidth]{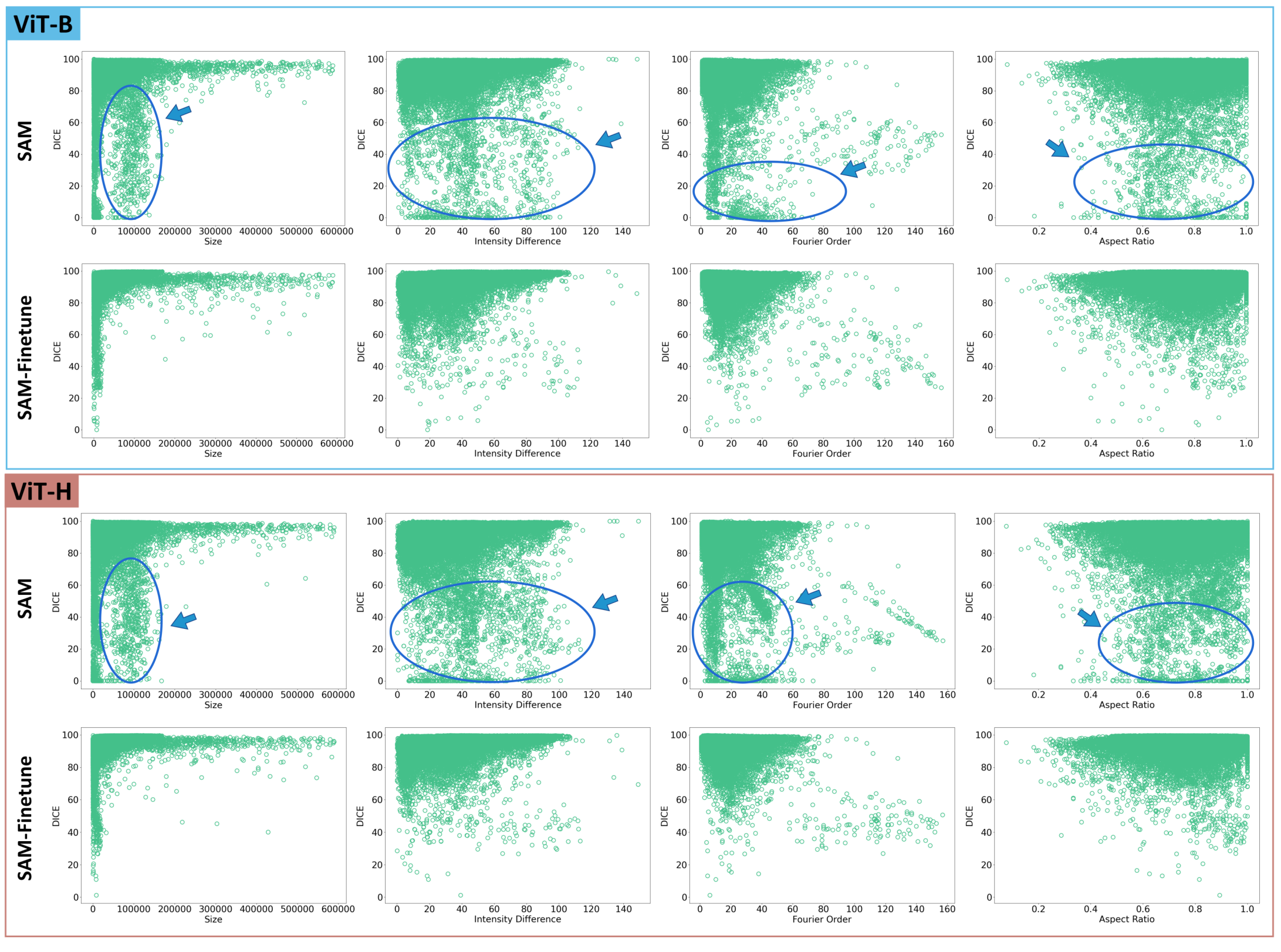}
	\caption{Trend analysis of DICE under different attributes (\textit{ViT-B} and \textit{ViT-H} with box prompt, $S_5$). The blue circles show the most obvious changes.}
	\label{fig:fine_tune_before_after}
\end{figure*}

\begin{figure*}[!ht]
	\centering
	\includegraphics[width=1.0\linewidth]{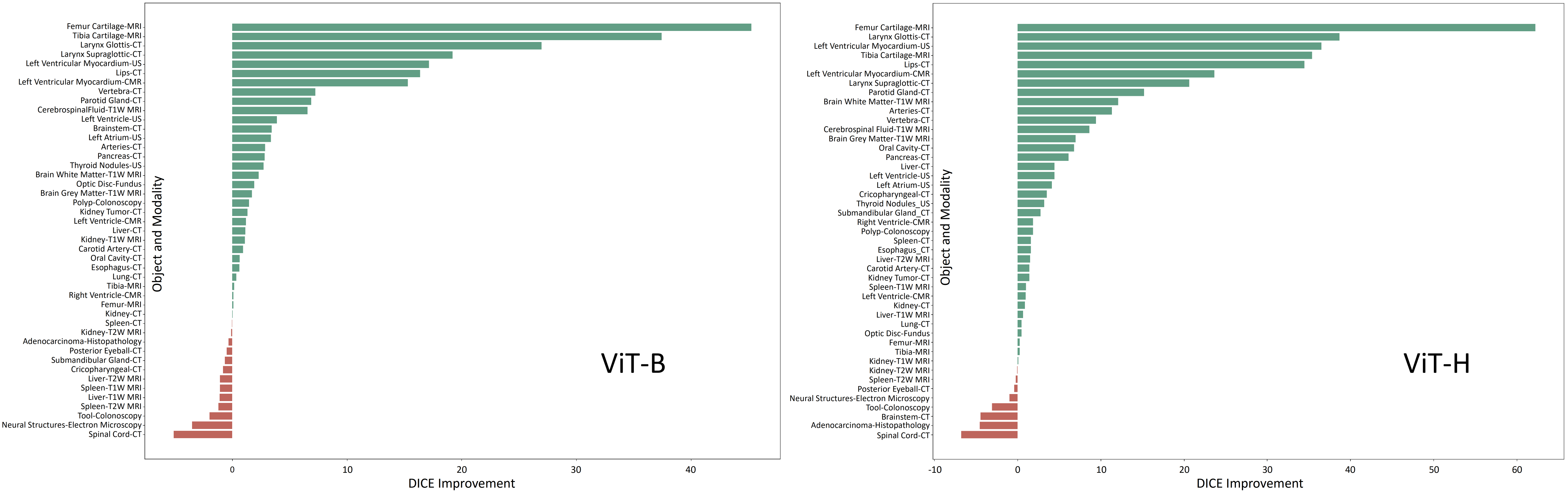}
	\caption{DICE improvement after finetuning, including SAM with \textit{ViT-B} and \textit{ViT-H}.}
	\label{fig:fine_tune}
\end{figure*}

\subsection{Comparison between SAM and Interactive Methods}
In the previous sections, we input all the prompts to the SAM's prompt encoder once for a fair comparison of its one-round performance.
To mimic the real-life interactive segmentation procedures, we performed multi-round SAM.
The point selection strategy is similar to the common interactive methods.
Specifically, SAM first clicks on the center of the target, then the remaining clicks are based on false negative (FN) and false positive (FP) regions.
We then compared SAM with two different strong interaction segmentation approaches, i.e., FocalClick~\citep{chen2022focalclick} and SimpleClick~\citep{liu2022simpleclick}. 
Both of them are pretrained on the same number of images as SAM was.

We selected 10 typical organs/tumors, covering various modalities, shapes, sizes, and intensity distributions.  
Experimental results are shown in Fig.~\ref{fig:interaction}.
Based on the DICE results, our conclusion is that:
1) SAM outperformed FocalClick and SimpleClick in the first interaction with a single point;
2) As iterations progressed, SAM's performance increased slowly, or even declined, while the performance of the interactive method could be improved steadily;
3) Using 10 points, SAM performed worse than the interactive methods.
Similar results can be found in the recently published MedIA paper~\citep{mazurowski2023segment}.
We consider that the current SAM's point-based multi-round iteration capability is weak on medical images.
Future work should optimize iterative training strategies when training SAM or finetune it to enhance its ability to iterate multiple rounds~\citep{cheng2023sam}.

\subsection{Task-specific Refinement for SAM}
SAM's weak ability perception on most medical images/tasks is mainly due to the lack of training data. 
The training dataset of SAM, i.e., SA-1B\footnote{\url{https://ai.meta.com/datasets/segment-anything/}} contains 11 million photos, including natural locations, objects, and scenes, but without any medical images.
Natural images are commonly different from medical images because they have color-encoding, relatively clear definitions and boundaries of objects, easier-to-distinguish foreground (objects) and background (non-objects), and relatively balanced size.
However, most medical images are gray-scale, with unclear and complex object boundaries, similar back- and fore-ground, and wide-range image size (especially containing some very small objects).

Thus, we finetuned SAM using part of the \textit{\ourdata} to improve SAM's perception of medical objects.
Specifically, we considered 45 common and typical objects for finetuning the SAM.
Inspired by \citet{ma2023segment}, we only considered finetuning the SAM using box prompts.
We fixed the image encoder to minimize computation costs and also kept the prompt encoder frozen because of its powerful capacity for encoding box positional information. 
Thus, only the parameters in the mask decoder were adjusted during finetuning.
We set the total epoch as 20, and the learning rate and batch size are 1e-4 and 2.

The results demonstrate a general improvement in segmentation performance after finetuning for both \textit{ViT-B} and \textit{ViT-H} models, as shown in Fig.~\ref{fig:fine_tune_before_after} and Fig.~\ref{fig:fine_tune}.
Fig.~\ref{fig:fine_tune_before_after} illustrates a shift towards higher DICE values across different correlating factors, indicating improved overall performance.
Specifically, for \textit{ViT-B}, 32 out of 45 objects exhibit performance enhancement, while \textit{ViT-H} shows improvement in 37 out of 45 objects. 
This can prove the strong learning ability of \textit{ViT-H}, because it has almost 7 times the parameters as \textit{ViT-B} (636M \textit{vs.} 91M).
The performance of objects with small numbers, RGB color coding, etc., decreases.
This reminds us that we may need to make more careful designs for task-specific finetuning.
We will release all the finetuned models and source codes.

\section{Conclusion}
In this study, we comprehensively evaluated the SAM for the segmentation of a large medical image dataset. Based on the aforementioned empirical analyses, our conclusions are as follows: 
1) SAM showed remarkable performance in some specific objects but was unstable, imperfect, or even totally failed in other situations.
2) SAM with the large ViT-H showed better overall performance than that with the small ViT-B. 
3) SAM performed better with manual hints, especially box, than the \textit{Everything} mode.
4) SAM could help human annotation with high labeling quality and less time.
5) SAM is sensitive to the randomness in the center point and tight box prompts, and may suffer from a serious performance drop.
6) SAM performed better than interactive methods with one or a few points, but will be outpaced as the number of points increases. 
7) SAM's performance correlated to different factors, including boundary complexity, etc.
8) Finetuning the SAM on specific medical tasks could improve its average DICE performance by 4.39\% and 6.68\% for \textit{ViT-B} and \textit{ViT-H}, respectively.
Finally, we believe that, although SAM has the potential to become a general MIS model, its performance in the MIS task is not stable at present.
We hope that this report will help readers and the community better understand SAM's segmentation performance in medical images and ultimately facilitate the development of a new generation of MIS foundation models.

\section{Discussion}

We will focus on discussing the potential future directions of SAM, and we hope these can inspire the readers to some extent.

\textit{How is semantics obtained from SAM when there is no GT?}
The current SAM only has the ability to perceive objects but cannot analyze specific categories of objects.
Recently, several studies have explored to address this problem, and one of them equipped the SAM with a CLIP model\footnote{\url{https://github.com/Curt-Park/segment-anything-with-clip}}.
Specifically, SAM will first provide the region proposals, and the region patch will be cropped from the original image.
Then, the cropped patch will be input to CLIP for object classification.
Another solution is to combine the SAM with an Open-Vocabulary Object Detection (OVOD) model, e.g., Grounding DINO with SAM (Grounded-SAM\footnote{\url{https://github.com/IDEA-Research/Grounded-Segment-Anything}}). 
In this pipeline, the OVOD model can detect the bounding boxes of objects with classification results. Then, SAM will take the box region as input and output the segmentation result.
Recently, semantic-SAM has been proposed to segment and recognize anything in natural images~\citep{li2023semantic}.
All the previous explorations are based on natural images. Thus, it may be interesting to develop medical SAM with semantic awareness.
However, it is challenging because medical objects in the open scene have varied and complex shapes, a wide variety of types, and many similar subclasses (tumors of different grades, etc.).

\textit{SAM \textit{vs.} traditional segmentation methods?}
Finetuning SAM with limited medical data can outperform task-specific traditional segmentation methods.
This has been validated in several recently published studies.
Med-SAM has proven that finetuning the 2D SAM can achieve superior performance to specialist Unet models in most cases~\citep{ma2023segment}.
3D modality-agnostic SAM (MA-SAM) has validated that finetuning SAM with 3D adapters can outperform traditional SOTA 3D nn-Unet, even without any prompts~\citep{chen2023masam}.
It also sheds some light on the medical image segmentation community, suggesting that perhaps finetuning the fundamental segmentation model will perform better than training a traditional segmentation model from scratch.
However, there are still some problems with SAM, including model robustness to different prompt noises and multi-round interaction ability.


\textit{2D or 3D SAM?}
For medical data, variability in imaging modalities (2D/videos/3D/4D) may make the design of general models complex.
Compared to videos/3D/4D images (CT/MRI, etc.), 2D is more foundational and common in medical data.
Thus, it is more practical to build a 2D model that can process all types of data consistently, as video/3D/4D data can be transferred to a series of 2D slices~\citep{ma2023segment}.
The limited amount of 3D data (SAM: 11M images and 1B masks \textit{vs.} Ours: $<$10K volumes and $<$45K masks) may restrict the construction of 3D fundamental segmentation models, especially if training from scratch is required.
To break the limitation of data, we will explore how to synthesize more high-fidelity 3D data and build a strong fundamental model for medical image segmentation.

\textit{SAM boosts large-scale medical annotation?}
Developing robust and effective deep learning-based medical segmentation models highly requires large-scale and fully-labeled datasets.
This is very challenging for current schemes based on manual annotation by experts.
As introduced in~\citep{qu2023annotating}, annotating 8,448 CT volumes with 9 anatomical structures and 3.2 million slices requires roughly 30.8 years for one experienced expert.
They shorten the annotation time to three weeks with the help of multiple pre-trained segmentation models for generating pseudo labels, and other useful strategies.
However, obtaining well-performing pre-trained models, especially with low false positives, is still very difficult.
Moreover, segmentation networks based on traditional deep learning cannot well support human-computer interaction, limiting its flexibility.
The occurrence of SAM with promptable segmentation brings hope to solve the challenges.
Our study also preliminarily verified that SAM can greatly shorten the annotation time and improve the annotation quality.
The greater the number of anatomical structures that need to be labeled, the more obvious the advantage of SAM's efficiency will be.
Notably, the design of the SAM paradigm has the potential to achieve universal segmentation.
It means that a single SAM network can be used to achieve the annotation of large-scale multi-modal, multi-class medical datasets, rather than using multiple task-specific models.
This is important for lightweight and efficient deployment of models in labeling software, e.g., MONAI Label~\citep{cardoso2022monai} and Pair annotation software package~\citep{liang2022sketch}.

\section{Acknowledgement}
The authors of this paper sincerely appreciate all the challenge organizers and owners for providing the public MIS datasets including 
AbdomenCT-1K, ACDC, AMOS 2022, AutoLaparo, 
BrainPTM 2021, BraTS20, 
CAMUS, CHAOS, CHASE-DB1, Chest CT segmentation, CRAG, crossMoDA, CVC-ClinicDB, 
DRIVE, 
EndoTect 2020, EPFL-EM, ETIS-Larib Polyp DB, 
FeTA, 
HaN-Seg, 
I2CVB, iChallenge-AMD, iChallenge-PALM, IDRiD 2018, iSeg 2019, ISIC 2018, IXI, 
KiPA22, KiTS19, KiTS21, Kvasir-Instrumen, Kvasir-SEG, 
iVScar, LUNA16, 
M\&Ms, MALBCV-Abdomen, Montgomery County CXR Set, MRSpineSeg, MSD, 
NCI-ISBI 2013, CellSeg Challenge-NeurIPS 2022, 
PROMISE12, 
QUBIQ 2021, 
SIIM-ACR, SKI10, SLIVER07, ssTEM, STARE, 
TN-SCUI, 
TotalSegmentator, 
VerSe19$\&$VerSe20, 
Warwick-QU, WORD, and 
4C2021 C04 TLS01. 
We also thank Meta AI for releasing the source code of SAM publicly available.

This work was supported by the grant from National Natural Science Foundation of China (Nos. 62171290, 62101343), Shenzhen-Hong Kong Joint Research Program (No. SGDX20201103095613036), and Shenzhen Science and Technology Innovations Committee (No. 20200812143441001).

\section{Supplementary materials}
Comprehensive quantitative and qualitative results can be found at \url{https://github.com/yuhoo0302/Segment-Anything-Model-for-Medical-Images/blob/main/Supplementary_Materials.pdf}.
\bibliographystyle{model2-names.bst}\biboptions{authoryear}
\bibliography{references}

\end{document}